\documentclass[a4paper]{aa}  
\usepackage{graphicx}
\usepackage{txfonts}
\usepackage[dvips]{color}
\newcommand{\besancon}{Besan\c{c}on}

\newcommand{\zwi}{I\,Zw\,18}
\newcommand{\hi}{\ion{H}{i}}
\newcommand{\hii}{\ion{H}{ii}}

\begin{document}
\title{
Outer structure of  the Galactic warp and  flare: explaining the Canis
Major over-density
\thanks{Based on archival data collected at the European Southern Observatory,
La Silla, Chile.}  }
\subtitle{}

\author{Y. Momany\inst{1,2}
        \and 
        S. Zaggia\inst{3} 
        \and
        G. Gilmore\inst{4} 
        \and 
        G. Piotto\inst{2} 
        \and
        G. Carraro\inst{2,5}
        \and
        L. R. Bedin\inst{6}
        \and 
        F. De Angeli\inst{4}
      }

\offprints{Y. Momany}

\institute{INAF- Oss.  Astronomico di Padova,  Vicolo
dell'Osservatorio            5,                 I-35122        Padova,
Italy. \email{yazan.almomany@oapd.inaf.it}
\and Dip. di Astronomia, Universit\`a di Padova, Vicolo
dell'Osservatorio 2, I-35122 Padova, Italy. \email{momany,piotto@pd.astro.it}
\and INAF - Oss. Astronomico di Trieste, Via Tiepolo 11, 34131
Trieste, Italy.  \email{zaggia@oats.inaf.it}
\and Institute of Astronomy, University of Cambridge, Cambridge, CB3 OHA, UK. \email{gil,fda@ast.cam.ac.uk}
\and Andes Fellow, Departamento de Astronom\'ia, Universidad de Chile, Casilla 36-D,
Santiago, Chile \\  Astronomy Department, Yale  University, New Haven,
CT 06511, USA. \email{gcarraro@das.uchile.cl}
\and European Southern Observatory, Karl-Schwarzschild-Str. 2, 85748
Garching b. M\"unchen, Germany. \email{lbedin@eso.org}
}

\date{Received 22 August 2005; accepted 28 February 2006}

\abstract
{}%one
{In this paper we derive the structure of the Galactic stellar
warp and flare.} %two
{We  use 2MASS red clump and red giant stars, selected at
mean and fixed heliocentric distances of $R_{\odot}\simeq3$, $7$ and
$17$ kpc.}
{Our results  can be summarized as  follows:  
(i) a  clear stellar warp  signature  is derived  for the  3  selected
rings, proving that the warp starts already within the solar circle;
(ii)  the derived stellar  warp is consistent (both in amplitude
and  phase-angle)  with that for the  Galactic   interstellar dust and
neutral atomic hydrogen;
(iii) the consistency  and regularity of  the  stellar-gaseous warp is
traced out to about $R_{GC}\sim20$ kpc;
(iv) the Sun seems not to fall on the line of nodes. The stellar warp
phase-angle orientation ($\phi\sim15^{\circ}$) is close to the
orientation angle of the Galactic bar and this, most importantly,
produces an asymmetric warp for the inner $R_{\odot}\simeq3$ and $7$
kpc rings;
(v) a Northern/Southern warp symmetry is observed only for the ring at
$R_{\odot}\simeq17$ kpc, at which the dependency on $\phi$ is weakened;
(vi) treating  a mixture of  thin and  thick disk  stellar populations,  we
trace the variation with $R_{GC}$ of  the disk thickness (flaring) and
derive  an  almost  constant   scale-height ($\sim0.65$  kpc)   within
$R_{GC}\sim15$  kpc. Further out, the  disk flaring increase gradually
reaching a mean scale-height of $\sim1.5$ kpc at $R_{GC}\sim23$ kpc;
(vii) the derived outer disk  warping and flaring provide further
robust evidence that there is no disk radial
truncation at $R_{GC}\sim14$  kpc.}%three
{In  the particular case   of the  Canis Major  (CMa)  over-density we
confirm  its coincidence with the  Southern stellar maximum warp
occurring near  $l\sim240^{\circ}$  (for $R_{\odot}\simeq7$ kpc) which
brings down  the  Milky  Way   mid-plane  by  $\sim3^{\circ}$  in this
direction.
The regularity and consistency of the stellar, gaseous and dust warp
argues strongly against a recent merger scenario for Canis Major.
We present evidence to conclude that all observed parameters
(e.g.  number density, radial velocities, proper motion etc) of
CMa are consistent with it being a normal Milky Way outer-disk population,
thereby leaving no justification for more complex
interpretations of its origin.
The present analysis or outer disk structure does not provide a
conclusive test of the structure or origin of the Monoceros Ring.
Nevertheless, we show that a warped  flared Milky Way
contributes significantly at the locations of the Monoceros Ring.
Comparison of outer Milky Way \hi\ and CO properties with those of
other galaxies favors the suggestion that complex structures
close to planar in outer disks are common, and are a natural aspect of
warped and flaring disks.
}%four
{}%five
\keywords{Galaxy: structure -- Galaxy: formation -- galaxies:
interactions -- galaxies: dwarf -- galaxies: individual Canis Major --
galaxies: individual Monoceros Stream} 
%
%\authorrunning{Momany et al.}
%
\titlerunning{The Galactic stellar warp and flare}
\maketitle
\section{Introduction}
There was a revitalisation of interest in the outer disk of the
Milky   Way    with the     photometric   discovery  of   significant
over-densities of  F stars in the Sloan  Digital Sky Survey by Newberg
et al.  (\cite{newberg02}) towards the Galactic anti-center.
The reported over-densities defined a type of stellar {\em Ring}
structure,   relatively     confined  to      the   Galactic     plane
($|b|\le30^{\circ}$),   and  stretching   over   the sky     for
$\sim100^{\circ}$ in longitude centered in Monoceros.
A kinematic and spectroscopic study  of the Ring  structure by
Yanny et al.  (\cite{yanny03}) confirmed (i) a Galactocentric distance
of  18 and 20  kpc (respectively above and  below the Galactic plane);
(ii) an inconsistency of its velocity dispersion with typical Galactic
halo    and thick  disk   structures;  and   (iii)   a metallicity  of
[Fe/H]$\simeq-1.6$ consistent with halo populations.
In the context of continuing searches for outer structure
related to the Sgr dwarf, and other possible accretion events the
most natural   scenario    proposed by Yanny etal  for the   Monoceros   Ring
(Mon. Ring) was that it traced a remnant dwarf  satellite galaxy in the
late process of disruption (if not already dissolved).
Accumulating evidence supporting  the  existence  of  the Mon.    Ring
(Rocha-Pinto et al.  \cite{rocha03},  Crane  et al.    \cite{crane03},
Frinchaboy et  al.  \cite{frin04}) soon triggered a  search for a {\em
progenitor}, obviously carried out near the Galactic plane.

At the same time,  in an  alternative, more conservative
scenario, the existence of the Mon.  Ring was seen as the
consequence of perturbations in the outer disk caused by ancient
warps (Ibata et al. \cite{ibata03}).
Indeed,  the Mon.  Ring rotates  in a  prograde  orbit that  is almost
circularized, strongly suggestive of a disk origin.  
Nevertheless, simulations by Helmi (\cite{helmi03}) showed that
accretion   models, where   co-planar  streams can follow 
circular orbits, were indeed feasible.
This encouraged a search for a progenitor of the Ring: if there is
one (the Sagittarius dwarf) why not another.

In an analysis of  2MASS  data, Martin et    al.\ (\cite{martin04a})
assumed a   symmetric Galactic  vertical stellar distribution  around
$b=0^{\circ}$   and   searched for asymmetrical  differences  between
Northern and Southern star-counts. 
Among other features, they   pointed to an elliptical-shaped   stellar
over-density centered  at   $(l,b)=(240^{\circ},-7^{\circ})$.     They
interpreted  this   over-density as  the  core  of a  satellite galaxy
currently undergoing in-plane accretion, namely  the Canis Major (CMa)
dwarf spheroidal galaxy, the best Ring progenitor candidate.
%-

In Momany et al.  (\cite{momany04b}), we  highlighted the fact that in
the analyses  in Martin et  al.\ (\cite{martin04a})  and Bellazzini et
al.  (\cite{luna04}, first astro-ph version) the possible influence of
a Galactic  stellar warp on the  detection of a  vertically asymmetric
distribution so close to the Galactic plane was not considered.
Observationally, the warp is a bending of the Galactic plane
upwards in the first and second Galactic longitude quadrants
($0^{\circ}\le~l~\le180^{\circ}$) and downward in the third and fourth
quadrants ($180^{\circ}\le~l~\le360^{\circ}$).
We emphasise that the suggested reality of an outer stellar warp
was not a new proposal by us. Among other earlier studies, one of
particular relevance is that of Carney \& Seitzer (\cite{carney93})
who analysed the {\em ``Galaxy's own structure to obtain at
least a peak at the outer disk''}, i.e.  using certain lines of
sights, one can look away from the plane, reducing the foreground disk
signal as well as reddening-absorption-crowding and derive the age and
metallicity of the outer warped disk.
Analyzing the color-magnitude diagrams of  fields very near to the CMa
center [$(l,b)=(245^{\circ},-4^{\circ})$] Carney \& Seitzer claimed to
have detected the main  sequence and turnoff region  of the {\em outer
Galactic disk}.
On the other hand, the analysis of Martin et al.\ (\cite{martin04a}) and
Bellazzini et al.  (\cite{luna04}) discounted the Galactic warp
in this zone, so that the stellar populations previously
identified as outer disk main sequence, were now proposed as an
un-expected stellar population.  Further hints of a 
star-count anomaly at $l=240^{\circ}$  are found in Alard
(\cite{alard00}) who, again, associated the {\em ``strong asymmetry''}
in this region with the Galactic stellar warp.

In Momany et al.  (\cite{momany04b}) we concluded that the CMa
over-density can be fully accounted for if the Galactic disk (at
$l=240^{\circ}$) is $2^{\circ}$ displaced/warped below the mid-plane;
i.e.  the symmetry axis for this region is at $b=-2^{\circ}$ and not
$b=0^{\circ}$.
In response to our analysis, Martin et al.  (\cite{martin04b})
presented radial velocities and argued that the Galactic stellar warp
(location and amplitude), { cannot} explain the CMa over-density.  In
particular, they argued that (i) a warp angle of $-2^{\circ}$ is { not
enough} to erase the CMa over-density; and (ii) the CMa over-density
is stronger in amplitude and located too far from the Southern
hemisphere warp at $l=270^{\circ}$.
More recently, Rocha-Pinto et  al. (\cite{rocha05}) proposed that  the
amplitude  of the  CMa over-density  is  small with respect to another
over-density, this time located in Argo ($l\sim290^{\circ}$).  In this
later analysis, Rocha-Pinto et al.  view the  CMa over-density as  the
consequence of a dust  extinction window aligned  with a  maximum warp
location  at $l=245^{\circ}$ (as seen  in L{\' o}pez-Corredoira et al.
\cite{lopez02}).  Confusingly, Martin  et  al.  (\cite{martin04b}) and
Bellazzini   et al.   (\cite{luna05})   use  the  same   source  (L{\'
o}pez-Corredoira et al.  \cite{lopez02})  to  argue that  the  maximum
warp location is at $l=270^{\circ}$.

In  order to clarify  this situation, it is  timely to re-evaluate the
detailed properties of   the Galactic stellar   warp, specifically the
location and amplitude of its maximum.
Most recently,   Conn et al.   (\cite{conn05}) presented  a wide-field
survey of   the Mon.  Ring  and emphasised  that the presence  of Ring
streams  above and below the Galactic  plane  argue against a Galactic
origin of  the  Mon.  Ring.   Moreover,  they suggested  that positive
detections of the Mon.   Ring below the  plane may also be  correlated
with the Triangulum-Andromeda (TriAnd) structure (Majewski et al.
\cite{majewski04} and   Rocha-Pinto et al.   \cite{rocha05}).  In this
scenario, although the Mon.  Ring and  TriAnd structure are located at
different distances, the TriAnd structure  could be the distant arm of
a multiply-wrapped tidal stream.

In the main part of this  paper we use  2MASS  data to { derive}
and trace the signature of the Galactic stellar  warp as a function of
Galactic longitude.
Our working hypothesis is simple: had the CMa over-density been due to
an  extra-Galactic  accretion,  one would expect   it  to appear as  a
distortion on  top of a  large-scale  structure, that is  the Galactic
stellar  warp.  As  we shall  demonstrate, CMa  as  an over-density is
easily accounted for as being the maximum Southern stellar warp. This,
in our  opinion, allows Occam's  Razor to  indicate a clear preference
for a   Galactic structural origin of the   CMa over-density.  We then
discuss evidence which we believe associates the Mon.  stream with the
warped and flaring Galactic disk.
In several Appendices we briefly review and comment on important
aspects of the observational properties of CMa and the Mon Ring.

\section{The Galactic Stellar Warp}

In Bellazzini et al.   (\cite{luna05}) the authors  commented on
how the different adopted  parametrization of the warp  may lead
to some confusion.  Indeed, the warp has been derived from: (1) the mean
latitude of the adopted tracer as a function of longitude (Djorgovski
\& Sosin \cite{djor89}); (2) the ratio of star counts in Northern and
Southern hemispheres as a function of longitude (L{\' o}pez-Corredoira
et al.  \cite{lopez02}); and (3) the latitude  of peak brightness as a
function of longitude (Freudenreich et al. \cite{freud94}).  There has
been some disagreement  about these determinations.  Bellazzini et al.
(\cite{luna05}) argued that   the Momany etal  parametrization  of the
warp  was ``{\em   not   a fair   description}''  of  the  South/North
over-densities.  In the following we will firstly expand on our method
of tracing the  Galactic warp and then  secondly  we will compare  our
results with  those derived  by  other methods.   This comparison show
excellent consistency.

In Momany et al.  (\cite{momany04b}, Fig.~2) we extracted dereddened
2MASS M-giants ($0.85\le(J-K)_{\circ}\le1.3$) within an oblique box
surrounding the CMa red giant branch  in the CMD, and falling in
a strip between $-20^{\circ}\le~b~\le20^{\circ}$ and
$235^{\circ}\le~l~\le245^{\circ}$.
Initially assuming reflection symmetry around $b=0^{\circ}$
(warp angle $w=0^{\circ}$) we performed Northern and Southern star
counts, binned in $0.5^{\circ}$.
This showed clearly ({\it upper panels of the same Figure}) that
by assuming a symmetry around $b=0^{\circ}$, as done in Martin
et al.  (\cite{martin04a}), one can  recover the identified (CMa)
over-density.
Successively     the   symmetry axis  was varied within
the range $|b|\le5^{\circ}$, in steps of $0.1^{\circ}$.
For each  step  (i.e.   each warp  angle)  the Northern   and Southern
latitude profiles    were  folded  and a  reduced    $\chi^{2}$
defined as:
 
\begin{equation}
\chi^2 = \sum [(y_{North}-y_{South})^2/(y_{North}+y_{South})] 
\label{e_sum}
\end{equation}
%
%-------------------------------------------------------------
%
was employed to determine   the warp angle which  minimized  the
differences between the two profiles ($y$ is the Log(N) of star counts
computed in the latitude profiles, see Fig.~\ref{f_fig3}).
A    symmetry of    around   $b=-2^{\circ}$   (i.e   warp  angle
$w=-2^{\circ}$)   almost      completely   canceled  the   (CMa)
over-density.  This warp angle has been {\em derived} [not modeled
as argued  in Martin et  al.  (\cite{martin04a}) and Bellazzini et al.
(\cite{luna04})] and agrees with the angle derived for the
gas warp by Freudenreich (\cite{freud94}).
It is important to note that this  result is obtained independently of
the   adopted   dereddening   method:    (i)  pure Schlegel     et al.
(\cite{schl98}) values;  or (ii) modified   with the Bonifacio  et al.
(\cite{boni00})   correction    [see  Am{\^ o}res    \&   L{\'  e}pine
(\cite{amores05}) for recent confirmation of errors in the Schlegel et
al. values].

Thus, we are able to measure the latitude angle  at which the latitude
profiles crossing the Galactic disk are symmetric; i.e. the warp-angle
or the mid-plane  of the warped disk.   In this section we apply  this
method         for  the     entire      2MASS      catalog      within
$-20^{\circ}\le~b~\le20^{\circ}$, and derive  the global Galactic warp
as  a function of longitude.   Most importantly, the warp signature is
traced by  means of different stellar sources  [red clump (RC) and red
giants (RGB)] at  different distances.  This is particularly important
because we want to investigate  the impact of the tracer contamination
and its distance on the derived warp signature.

To emphasize the  particular   importance  of the contamination    and
distance    properties  of  the  assumed   tracer   (and the intrinsic
difficulty for  similar investigations)  we will briefly comment
on  two  recent   studies  probing the  Galactic warp, namely
Yusifov   (\cite{yusifov04})   and   L{\'    o}pez-Corredoira et   al.
(\cite{lopez02}).

\subsection{The Yusifov stellar warp model}
\label{s_yusifov}
%

%------------------------------------------one column figure
\begin{figure}[h]
\centering \includegraphics[width=7cm,height=5cm]{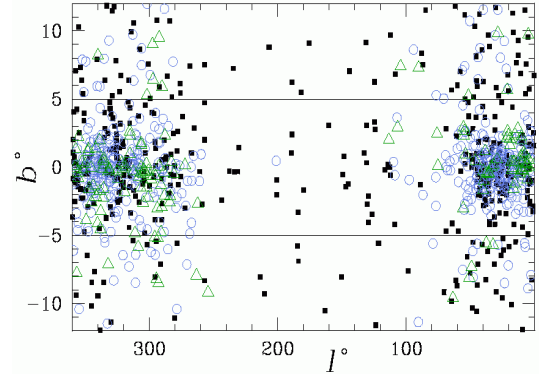}
\caption{The distribution pulsars in the Manchester et
al. (\cite{manchester05}) catalog. Filled squares and
open circles are pulsars with Galactocentric distances between 1 and 5
kpc, and  5 and 10   kpc respectively. Open  triangles  are pulsars at
$R_{GC}\ge10$ kpc.}
\label{f_fig1}
\end{figure}

In  his analysis, Yusifov  used   the asymmetric distribution of  1412
Galactic pulsars (from the Manchester et al.
\cite{manchester05} catalog)   to trace the  Galactic  stellar warp
and flare   (the    increase  in  scale-height  as   a   function   of
Galactocentric  distance).  Yusifov  limited  his analysis to  pulsars
within  $|b|\le5^{\circ}$ whose distances  are  $\ge 1$ kpc. He
calculated  the ratio of  the  cumulative number  of pulsars above  and
below the Galactic   plane as a function   of  Galactic longitude,  and
so derived a warp model.
However, as    seen  in  Fig.~\ref{f_fig1},  the Manchester    et  al.
(\cite{manchester05}) catalog  {\em has  only  1 pulsar} at $R>5$  kpc
between $200^{\circ}\le~l~\le270^{\circ}$.
Clearly, the pulsar catalog is already strongly incomplete at $R\sim5$
kpc in the outer  Milky Way.  Although one might  doubt how these data
can reliably predict the number density of stars  at CMa distances, it
is true that the Yusifov warp model  predictions have turned out to be
comparable  with other studies based  on more complete samples, and we
therefore will compare our results with this model.

\subsection{The L{\' o}pez-Corredoira et al. stellar warp model}

The  L{\' o}pez-Corredoira   et al.  2MASS-based   investigation was a
major attempt at studying the Galactic  stellar warp and flare.  There
are, however, a  few points that must be  kept in mind regarding their
analysis.  Firstly, 2MASS  was  not complete  at  the time that  their
analysis was concluded, so  that they were   able to analyse  only 820
lines of sight, each  of area between  $0.5$ and $1.0$ square degrees,
restricted to $|b|=\pm3^{\circ},\pm6^{\circ},\pm9^{\circ}$.
This is particularly  important in understanding the determination (cf
Martin et al.  \cite{martin04b} and Bellazzini  et al.  \cite{luna04},
\cite{luna05}]   of    the   maximum of    the     stellar   warp   at
$l\sim270^{\circ}$, a result often cited from the L{\'o}pez-Corredoira
et al. paper.
Indeed, the   area around $l\sim270^{\circ}$  was  missing in the L{\'
o}pez-Corredoira  et al.  study,  and { nowhere in  that article is it
stated   that the stellar   warp   maximum is at  $l\sim270^{\circ}$}.
Indeed the formula describing  the L{\'o}pez-Corredoira et  al.  warp
model at CMa distances shows that the maximum stellar  warp is near to
$l\sim240^{\circ}$ (see Section~\ref{s_rgb_warp}).
The value of  $l\sim240^{\circ}$ for the maximum  Galactic warp is  in
fact found in other studies, e.g.  Freudenreich (\cite{freud94}, their
Fig.~3)  studying  the    gaseous   warp, and   Djorgovski  \&   Sosin
(\cite{djor89},  their   Fig.~1)  analyzing IRAS   sources\footnote{In
particular note  a  clear warp  excess  around $l\sim240^{\circ}$ with
respect to their fitted function.}.

%------------------------------------------one column figure
\begin{figure}[h]
\centering \includegraphics[width=8.5cm,height=10.5cm]{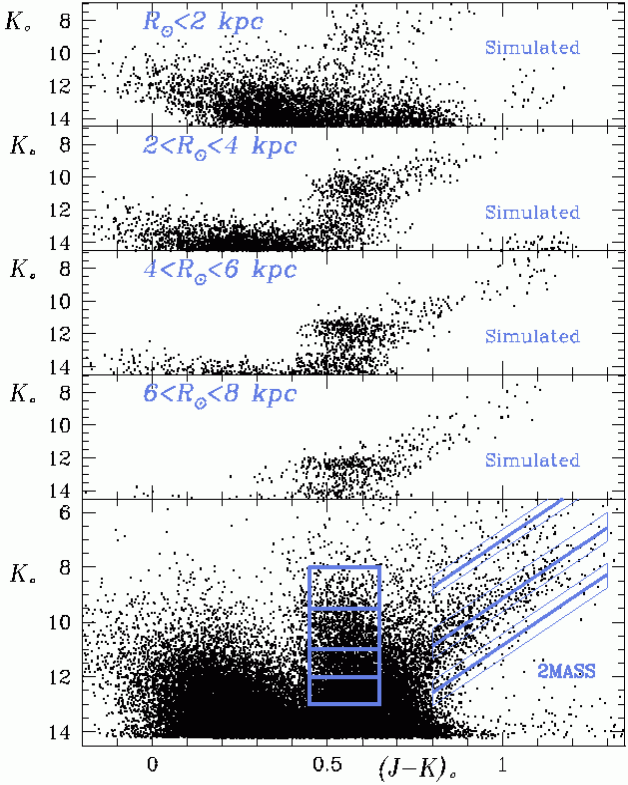}
\caption{The upper 4 panels display a \besancon\ $1^{\circ}$ square
area simulated color-magnitude diagram around CMa
($l,b$)$=(240^{\circ},-7^{\circ})$.   These show the expected
different contributions of stars as a function of their distance.  The
lowest panel displays  data from 2MASS between
$238^{\circ}\le~l~\le242^{\circ}$ and $-8^{\circ}\le~b~\le-6^{\circ}$.
The outlined vertical boxes mark 4 distance intervals as selected from
the red clump  distribution in simulated diagrams.
The 3 oblique boxes trace 3 samples of  red giant stars extracted
at three different heliocentric distances.
The  thick line is the Majewski et al. calibration of the 
Sagittarius dwarf RGB, shifted to $2.8$, $7.3$ and $16.6$ kpc.}
\label{f_fig2}
\end{figure}
%-------------------------------------------------------------

Second, L{\' o}pez-Corredoira et al.   derive the ratio of Northern to
Southern star-counts,  $R_{RC}=N_{North}/N_{South}$ for red clump stars with
$K_{\circ}\le 14.0$ (see their Fig.~15).
To help us evaluate the   impact of contamination when using  RC
stars as warp-tracers, in Fig.~\ref{f_fig2} we use \besancon\ (Robin
et al. \cite{robin03}) simulated  color-magnitude diagrams  around CMa
and plot the data as a function of their distance.
Focusing our attention around ($J-K$)$_{\circ}\simeq0.6$, one sees
that the stellar populations at $\le2$ kpc which will contribute
to the final $R_{RC}^{K\le14.0}$ ratio include: (i) RC stars at
$8.0\le~K~\le10.0$, and most importantly, (ii) {\em dwarfs} at the
fainter magnitudes $K_{\circ}\ge11.5$.
At magnitudes around the RC of  CMa ($K\sim13.0$), we estimate a
contamination by local foreground dwarfs of about 25-30\%.
Thus,  if one is  interested  in the $R_{RC}$ ratio   at say, the  CMa
distance,  one must bear   in  mind that this    ratio is subject   to
potentially serious contamination by  RC and dwarf stars at  distances
closer than the CMa over-density.

%------------------------------------------one column figure
\begin{figure}[h]
\centering\includegraphics[width=9.cm,height=8.5cm]{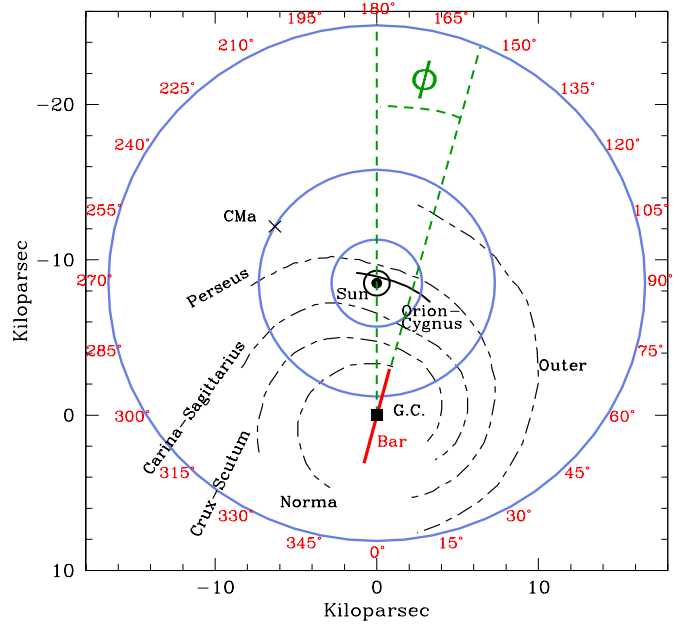}
\caption{A schematic view (from Drimmel \& Spergel  \cite{drimmel01}) of
the Milky  Way as seen from its  North pole showing the 4 spiral
arms  as   mapped by  \hii\ regions   and dust  (see also  Vall{\' e}e
\cite{vallee05} and Russeil
\cite{russeil03}).
The  Galactic center, the  Sun  and the names of   the spiral arms are
plotted.  Note  the presence of the  Local arm (Orion-Cygnus) close to
the Sun's  position  and  the  {\em outer arm}    that is also  called
Norma-Cygnus.
The 3 heliocentric  circles  define the  regions at  fixed $R_{\odot}$
where we  extract RGB samples  to trace  the Galactic  warp and flare.
The direction of the warp phase angle $\phi$, and longitude directions
are also plotted.
We have also added the  Galactic bar at  an orientation angle of
$14^{\circ}$  (following Freudenreich \cite{freud98}) with length
  3 kpc (following  Vall{\'e}e  \cite{vallee05}).  Note however
that  a  recent   Galactic  Legacy  Mid-Plane   Survey  Extraordinaire
(GLIMPSE) study by Benjamin et al.   (\cite{ben05}) report on a linear
Galactic bar with half-length of $4.4\pm0.5$ kpc.}
\label{f_rings}
\end{figure}

%-------------------------------------------------------------
%

\subsection{Tracing the stellar warp and flare with RC and RGB stars}
\label{s_RC_RGB}
We employ our method in tracing the  Galactic warp and flare using two
stellar  tracers: red clump and red  giant stars.  For both tracers we
repeat the analysis for different mean distances.
The   advantage in  using  two   stellar tracers   derives from  their
different contamination  status.  
With respect  to  RC stars,   red  giants are an   ideal instrument in
probing  the Galactic  warp and   flare since  these suffer less  {\it
external contamination} by nearby dwarfs.  At the same time red giants
projected at different distances do not overlap in the color-magnitude
diagram (i.e.  {\it no  internal contamination}) and this guarantees a
better distance separation.  Indeed, being the bright and evolved part
of the faint and un-detected main sequence stars,  red giants allow us
(already in the {\it  not so  deep} 2MASS catalog)  to probe  the most
distant, and almost entire, Milky Way disk populations.

%------------------------------------------one column figure
\begin{figure}[h]
\centering\includegraphics[width=7.5cm,height=5.5cm]{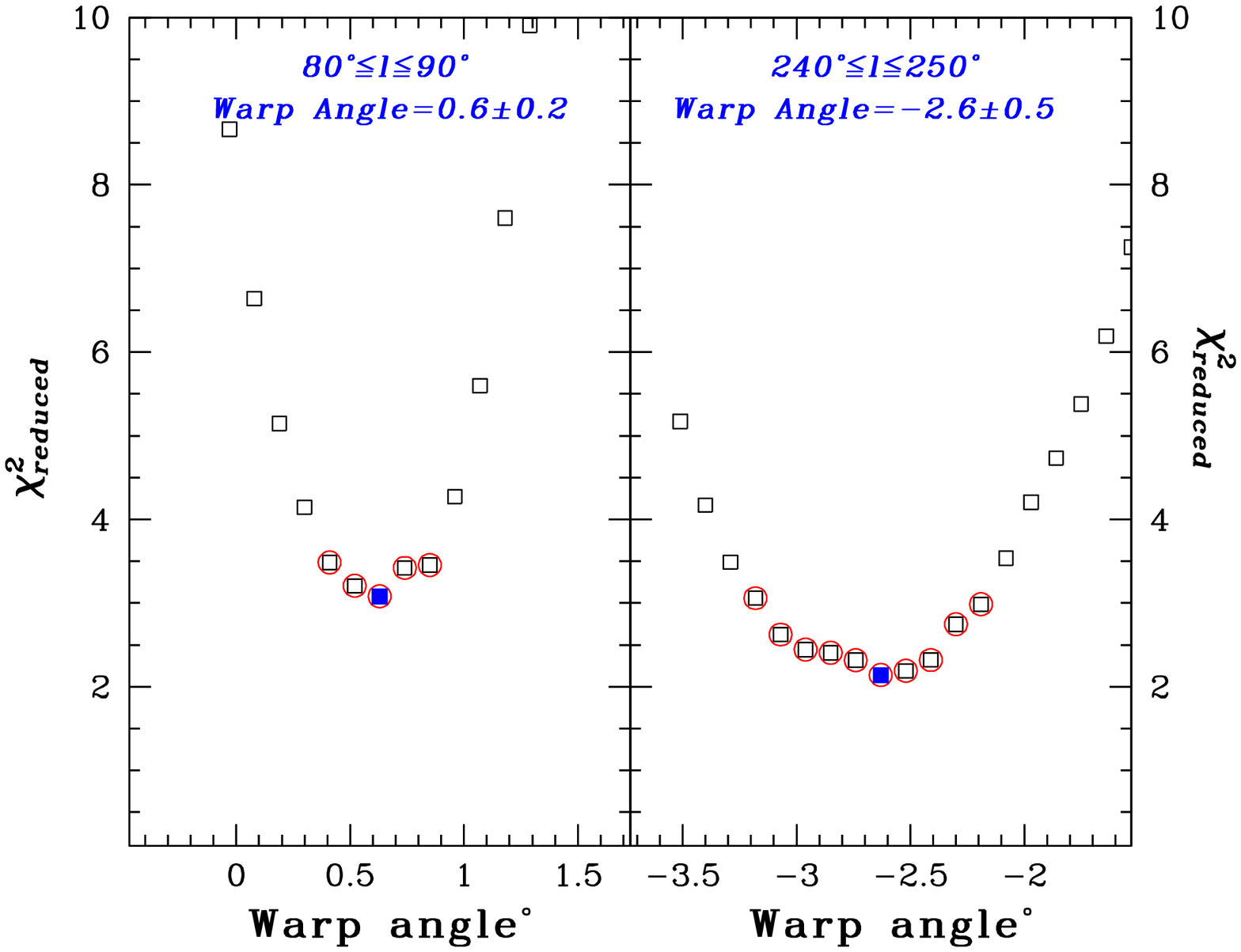}
\centering\includegraphics[width=8cm,height=9cm]{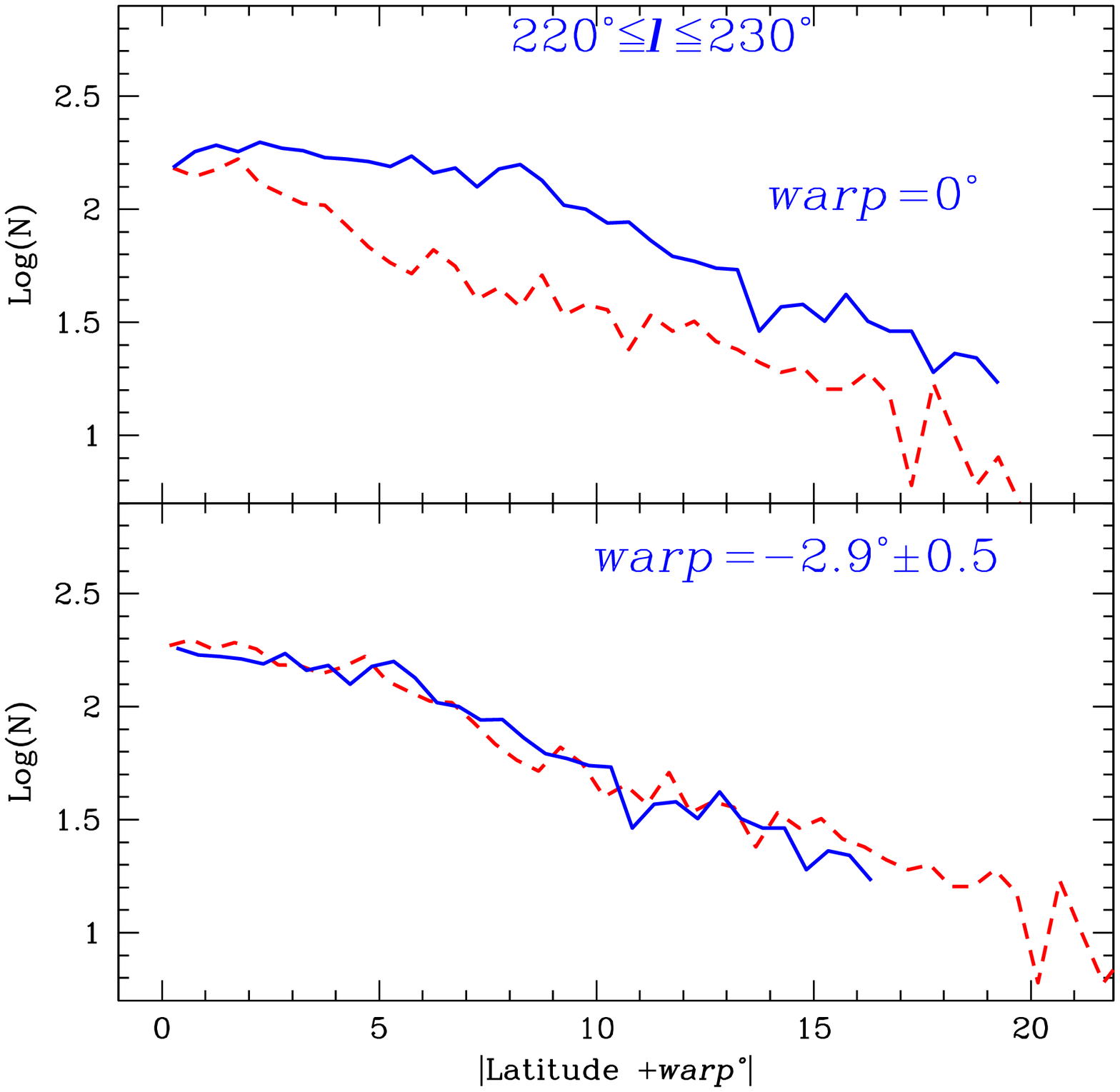}
\caption{The upper panels display two examples of the derivation of best 
warp angles and  their relative error  by means of a reduced $\chi^2$.
The lower panels  show the latitude profiles  for  a third field:  (i)
assuming no warp (upper panel) and; (ii)  having derived the best warp
angle (lower panel). The dashed and solid lines refer  to the Northern and
Southern latitude profiles, respectively.  All panels refer to the RGB
sample at $R_{\odot}=7.3$ kpc.}
\label{f_fig3}
\end{figure}
%-------------------------------------------------------------
%

%------------------------------------------one column figure
\begin{figure}[h]
\centering \includegraphics[width=7.5cm,height=5.5cm]{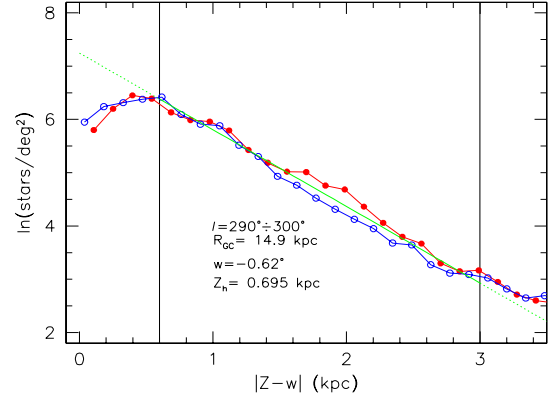}
\caption{An example of deriving the scale height ($Z_{\rm h}$) from the RGB sample
at  $R_{\odot}=16.6$ kpc.  The  filled circles show the Northern
hemisphere latitude profile  while the   open circles show   the
Southern profile.  Vertical lines delimit the  region where we apply a
linear fit (dotted line) to the data.}
\label{f_F1}
\end{figure}
%-------------------------------------------------------------

%
The  vertical  boxes  in  Fig.~\ref{f_fig2} set   4 heliocentric
distance  intervals  ($\le2$,  $2\div4$, $4\div6$   and  $6\div8$ kpc)
from which  we extract RC stars  and use their star-counts to derive
the warp and flare.

Similarly,   the 3  oblique  boxes trace   red   giant stars at  fixed
heliocentric   distances  between  $2.3\div3.2$,   $5.5\div9.1$,   and
$13.2\div20.0$   kpc.   Thus, the mean  assumed   distances of the RGB
samples are $2.8$, $7.3$ and $16.6$  kpc, with the intermediate sample
being centered on the CMa RGB.
To estimate the  distances to Milky Way M-giants we apply the
same method as in Martin et al.  (\cite{martin04a}), that first led to
the identification of the CMa over-density.
This is  done  by  using the  Sagittarius  dwarf RGB  as a  reference.
Assuming   a  distance   modulus  of  16.9  and   a   mean  [Fe/H]  of
$\sim-0.5$\footnote{A recent  study by Monaco et al. (\cite{monaco05})
report on high resolution UVES spectroscopy showing a mean metallicity
of [Fe/H]$=-0.41\pm0.20$.} Majewski et al. (\cite{majewski03}) derived
the  following    calibration      of      the    Sagittarius     RGB:
$K=-8.650\times(J-K)_{\circ}+20.374$ (see      the   thick  line    in
Fig.~\ref{f_fig2}).
The  mean metallicity  of Sagittarius  can  be considered intermediate
between inner (more metal-rich) and outer  (more metal-poor) Milky Way
disk stars. Indeed, the  mean abundance of the  disk stars is known to
vary   between $-1.0\le$[Fe/H]$\le+0.3$    [see Bensby, Feltzing    \&
Lundstr{\"o}m (\cite{bensby04}) for differences between thin and thick
disk populations]

At the  same time we note that  the Majewski et al.  RGB calibration
would  reproduce  the entire range of  disk  metallicities if age is
allowed to vary. For example the Sagittarius RGB  would overlap with a
10 Gyr and [Fe/H]$=-0.7$ theoretical  isochrone (appropriate for outer
disk  populations?)     as well as  one with     4 Gyr and [Fe/H]$=-0.4$
(appropriate for inner disk populations?).
A systematic uncertainty in  estimating the distances of the  Galactic
M-giants is  therefore un-avoidable. Similarly,  contamination between
different populations  and uncertainties in reddening corrections will
cloud our analysis.
A schematic view  of the Milky  Way and of the  regions where we probe
the stellar warp and flare are shown in Fig.~\ref{f_rings}.

Examples of the   application of our method  are  shown in the   upper
horizontal  panels of Fig.~\ref{f_fig3},  reporting  the derivation of
the best warp angle for two fields  extracted from the 
$R_{\odot}=7.3$  kpc RGB sample.   We show two  cases: the first field
$80^{\circ}\le~l~\le90^{\circ}$  and     $|b|\le20^{\circ}$   is  more
populated  and   the warp angle (minimum    $\chi^{2}$, as reported in
equation \ref{e_sum}) is easier    to determine, having  a  relatively
small     $0.2^{\circ}$     error.    The          second       field,
$240^{\circ}\le~l~\le250^{\circ}$, is  less populated and the error is
higher reaching $0.5^{\circ}$.

The lower panels of Fig.~\ref{f_fig3}, show an example of deriving the
best       warp        angle  for     latitude      profiles   between
$220^{\circ}\le~l~\le230^{\circ}$ from  the $R_{\odot}=7.3$  kpc RGB
sample.
The  upper   panel shows the   Northern and  Southern latitude profile
assuming no warp, i.e. a symmetry axis around $b=0^{\circ}$. The lower
panel shows   the same two profiles  assuming  a symmetry  axis around
$b=-2.9^{\circ}$, as  derived when the  differences between the folded
profiles are minimized and the  symmetry angle is  allowed
to vary within $|b|\le5^{\circ}$.
One may note that for large warp angles the Southern latitude profile
becomes {\it shorter} than the Northern one.  This is due to the
warp angle becoming significant compared to the imposed angular
limit in   the  extraction; i.e.   $|b|\le20^{\circ}$.

Once the   warp angle has   been derived for   each  line of  sight we
proceed  in estimating the vertical  density profile  of the stellar
disk: the  scale-height  ($Z_h$). Our  intent is  to characterize  the
radial  trend of  the   stellar disk  scale-height.   A  {\em  disk
flaring}  is usually seen as   an increasing scale-height towards the
outer parts of the disk.
An example of the adopted procedure is shown in Figure~\ref{f_F1}.
Having derived  the  warp angle,  the $b$ angles  were transformed  in
linear $Z$ height   according to the distance    from the Sun  of  the
sample. Then the Northern and Southern profiles were overlapped and
the  star density  profile was  fitted by a  power law,  thus deriving
$Z_h$.
The fitting of the  vertical density profile was  made within 2 limits
in $Z$, so as to exclude: (i) the  highly obscured {\em inner} regions
within  $\pm2.0^{\circ}$ from the mid-plane  and;  (ii) the very  {\em
outer} regions where in cases of  high warp angles the latitude profile
of one  hemisphere  is {\em short} with  respect to   the other
hemisphere  (cf  the     latitude    profiles in  lower    panel   of
Fig.~\ref{f_fig3}).

The  vertical $Z$ scale-height was fitted  by a single exponential. We
have not tried  to use a more  complicated formula like  the $sech$ or
double   exponential   (see   for  example   the  analysis    by Alard
\cite{alard00}).
Thus, our analysis is aimed at measuring the order of magnitude of the
flare in the outer disk.  A  detailed parametrization of the flare (in
terms  of thin-thick disk  separation),  although very interesting, is
beyond the scope of this paper.
In particular,  we note  that for the   regions of most  interest (the
outer disk) any thin disk flaring  would act in   a way that increases  the
confusion between the thin and thick disk components.
This in-ability in separating  thin-thick disk components in the outer
regions is  exacerbated   by the unavailability  of  kinematic
all-sky   data, which might allow  a   separation   of  the   two
components. Thus,   for the outer  disk, one  is left with determining
only an approximate amplitude of any flare.
Figure~\ref{f_F1} shows an example of the vertical density profile fit
for     the RGB  sample   at  a    heliocentric  distance of  16.6 kpc
($R_{GC}\simeq14.9$ kpc).
The abscissa,  $|Z-w|$, indicates the linear  $Z$ height folded around
the mid-plane of  the disk as determined by  the warp  angle analysis.
The analysis of the radial trend of the  vertical scale height, $Z_h$,
for the RGB sample will be discussed in Section~5.

\subsection{Comparing DIRBE  integrated  surface  photometry with
2MASS star counts}
\label{s_dirbe}

In this  paper we  compare   the derived {\em stellar}
warp with  that obtained  for    other Galactic components,  including 
integrated light, neutral gas  and the interstellar dust.  
In regards  to  this, we mainly make  use  of the Freudenreich  et al.
(\cite{freud94}) study where the warp due to dust and integrated light
has been  derived  using the  Diffuse  Infrared Background  Experiment
(DIRBE) mapping of the Galactic plane (see also Vig, Ghosh
\& Ojha \cite{vig05}).

The results  of DIRBE   however come  from integrated    light surface
photometry which  sum contributions from composite stellar populations
at different  distances and luminosities,  and therefore these can 
be correctly compared with the  2MASS results only after a  discussion
of their relative weight.
The DIRBE $1.2 \mu$m and $2.2 \mu$m  spectral bands correspond roughly
to the  infrared $J$ and $K$ pass-bands.   It is important to note that
the DIRBE  near-infrared ($\lambda<5$$\mu$m) emission is  dominated by
{\em stellar disk  stars}    [whereas  in far-infrared bands     (e.g.
$240\mu$m) the emission by {\em interstellar dust} dominates]. 

To better understand which stellar populations  contribute most in the
DIRBE  emission maps, we used the  2MASS  star counts to integrate the
light coming from specific stellar populations. We limit this analysis
to only one line of sight (the 2MASS color-magnitude diagram presented
in  the lowest panel  of Fig.~\ref{f_fig2}),  and  expect  this to  be
representative of any other line of sight.
We  roughly disentangle three  main populations  in  this 2MASS  color
magnitude diagram:  (i)   MS objects  with $(J-K)<0.45$, mainly   main
sequence  and    blue super-giants   stars;    (ii)  RC   stars   with
$0.45<(J-K)<0.65$;  and (iii) RGB,    red super-giant  and  asymptotic
branch stars with $(J-K)>0.65$.
For each of  these three stellar   populations we summed  up the light
contribution coming from   the  single point sources   (stars) falling
within the above color intervals.

%------------------------------------------one column figure
\begin{figure}[h]
%\centering\includegraphics[width=8.5cm,height=3.5cm]{./psfiles/dirbe_cumj.ps}
%\centering\includegraphics[width=8.5cm,height=3.5cm]{./psfiles/dirbe_cumk.ps} 
%\centering\includegraphics[width=8.5cm,height=3.5cm]{./psfiles/dirbe_dist.ps}
\centering\includegraphics[width=8.5cm,height=8.5cm]{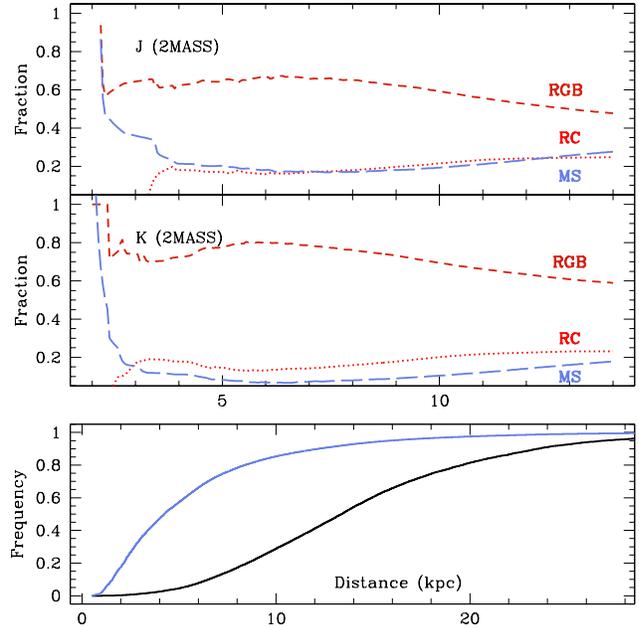}
\caption{The relative luminosity contribution of  the
three stellar populations in  the $J$ band  (upper panel) and $K$ band
(middle panel). The lower Figure shows the cumulative distributions of
the distances of RGB stars  without any luminosity-weighting along the
line of  sight (dark line).      The  grey line  shows instead     the
distance-weighted distribution.  }
\label{f_dirbe}      
\end{figure}

%-------------------------------------------------------------

In Fig.~\ref{f_dirbe} we  show the relative luminosity contribution of
the three stellar populations in  the $J$ band  (upper panel) and  $K$
band (middle panel). A limiting magnitude of $K=14.0$ has been applied
in  estimating the  cumulative  and relative  luminosity of  the three
different stellar populations.
The middle panel,  displaying the relative luminosity contribution  of
the three  populations to the total  luminosity, shows clearly that in
the case of  the $K$ band the total  light contribution of MS stars is
almost negligible in DIRBE, contributing only 15\% of the total light.
On the other hand, RGB stars contribute up to 60\% of the total light,
a level that goes up to 82\% once also the RC are summed together with
RGB stars.
Thus we conclude that the level  of contamination by  MS stars in $2.2
\mu$m ($\sim$$K$) DIRBE emission maps is fairly low and that {\it red}
populations dominate the integrated light.  Similar conclusions can be
drawn for the shortest DIRBE wavelength band at $1.2 \mu$m ($\sim$$J$,
upper panel): the  RGB and  RC populations  dominate the DIRBE  light,
contributing up to 75\% of the total light.

%------------------------------------------------------------------------------

We also  checked the  weighted distance  range at which  DIRBE is more
sensitive.  Considering that the   RGB   population is the    dominant
population in  both  $J$ and  $K$  bands we calculated  their distance
distribution using  the Majewski calibration   of the Sagittarius  red
giant branch (see Sect.~\ref{s_RC_RGB}).
The dark line   in  the lowest  plot of   Fig.~\ref{f_dirbe} shows the
cumulative   distributions of the distances  of  RGB stars without any
luminosity weighting along the line of  sight. The grey line shows the
cumulative distance distribution once  we weight the stellar  distance
with the stars light.

Clearly, the DIRBE integrated light is linearly sensitive to distances
within $  \sim9$ kpc (at  least in this  direction $l\sim240^{\circ}$)
where the light contribution reaches $80\%$ of the total.
This result, based  on RGB stars,  should not change considerably once
we add  the  contribution of (i)  the  RC stars  which weight  more to
larger distances;  and (ii) the MS  stars which weight more to shorter
distances.     Thus, being   of  the    same  size,  the two  opposite
contributions of the MS and RC stars will to first order cancel out.

In conclusion,   with  the help of    the 2MASS star  counts   we have
identified which stellar populations are  the main contributers to the
DIRBE  luminosity, and at   what distances these  contribute the most.
The results  are that the RGB stars  are the  dominant contributors of
the  $J$ and $K$  DIRBE integrated  light, within  $\sim9$ kpc.   This
allows us to perform  a fruitful comparison of  the DIRBE warp results
with those obtained from our 2MASS analysis.
%-----------------------------

%
%------------------------------------------one column figure
\begin{figure}[h]
\centering\includegraphics[width=8.5cm,height=8.5cm]{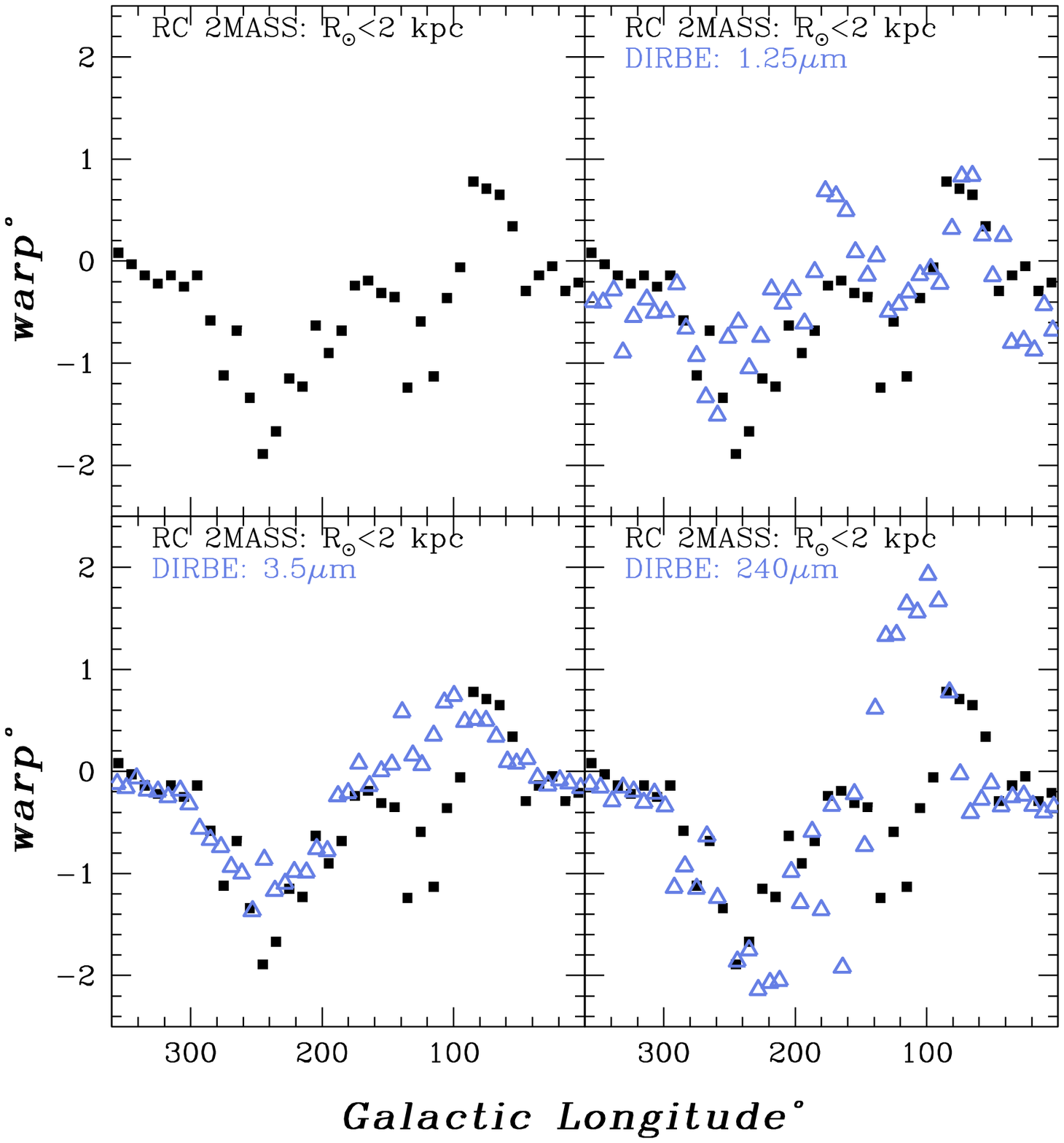}
%\centering\includegraphics[width=8.5cm,height=8.5cm]{./psfiles/fig6.ps}
\caption{The stellar warp as derived from RC
stars at $R_{\odot}\le2$ kpc  (filled squares).  The results are
compared with  the  {\em latitude  of peak  brightness} obtained  from
DIRBE  at wavelengths 1.2,  3.5 and 240  $\mu$m by Freudenreich et al.
(\cite{freud94}, grey open triangles).}
\label{f_fig4a}
\end{figure}
%-------------------------------------------------------------

\section{The stellar warp as traced by RC stars}
\label{s_rc}
%------------------------------------------one column figure
\begin{figure}[h]
\centering\includegraphics[width=8.5cm,height=8.5cm]{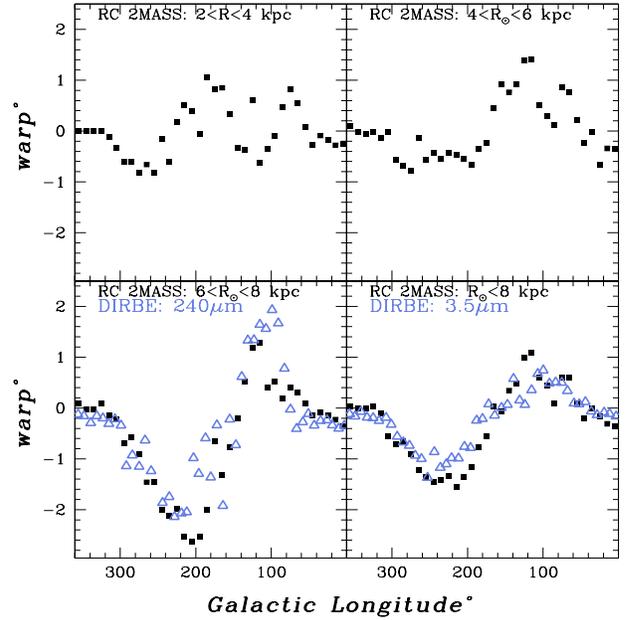}
\caption{The stellar warp as derived from RC stars for 4 distance
intervals (dark filled squares).  The results are compared with
the {\em latitude of peak brightness} obtained from DIRBE at
wavelengths 3.5 and 240 $\mu$m by Freudenreich et al.
(\cite{freud94}, grey open triangles).}
\label{f_fig4b}
\end{figure}
%-------------------------------------------------------------

In  this section we  present the derived stellar  warp as traced by RC
stars.  We  anticipate two  obvious   points: (i) the warp   amplitude
depends on the employed  wavelength; shorter wavelength investigations
probe nearer areas and therefore  may underestimate the warp amplitude
if it  increases with distance;  and (ii)  the warp amplitude  derived
depends on the contamination of the employed stellar tracer: inclusion
of  nearby stars in samples of   more distant stars will underestimate
the warp amplitude at larger distances.   Overall, red clump stars are
not the ideal  tracers of the warp  at  larger distances,  and we will
show this using Fig.~\ref{f_fig4a} and ~\ref{f_fig4b}.

The  upper left panel  of Fig.~\ref{f_fig4a} displays the stellar warp
as  derived using RC stars at   $R_{\odot}\le~2$ kpc.  At first sight,
the presence of  a global stellar  warp can be  debatable, as it contrasts
with the neat  {\em sinusoidal} function  seen in  Djorgovski \& Sosin
(\cite{djor89}) for IRAS sources.
It remains true however that the  Southern stellar warp is  distinguishable
between  $190^{\circ}\div290^{\circ}$.    This brings     the Galactic
mid-plane  $\sim2^{\circ}$  down  at  around  $l\sim250^{\circ}$,  and
demonstrates that the stellar warp is detectable  already in the solar
circle.
Problems  arise when  searching  for a  global  stellar  warp  in  the
Northern hemisphere. A sudden  drop at around $l\sim120^{\circ}$ seems
to interrupt a global warp signature.
To  understand the  absence of  a clear  stellar warp  in the Northern
hemisphere  we make a  comparison   with  the  {\it latitude  of  peak
brightness} as derived in  Freudenreich et al.  (\cite{freud94}) using
DIRBE mapping of the Galactic plane.
Overall, the DIRBE   data-points at 1.2$\mu$m  reproduce the global
features derived from nearby $R_{\odot}\le~2$ kpc RC stars.
In particular both data-sets  show an early drop at $l\sim120^{\circ}$
demonstrating: (i) the  consistency of the DIRBE-2MASS comparison; and
(ii) the presence of a localized nearby structure.
Indeed, the   drop  in 2MASS  and DIRBE  traces  the  presence  of the
Orion-Cygnus segment (called Local Arm in Russeil
\cite{russeil03}).

The  lower  left  panel  of Fig.~\ref{f_fig4a}   over-plots  the DIRBE
data-points at  3.5$\mu$m.  As anticipated before,  at $\lambda<5\mu$m
DIRBE emission is still dominated by disk stars. However, as one moves
to longer wavelengths DIRBE data-points sample more distant structures
that were obscured   at shorter  wavelengths.   This  explains why  at
3.5$\mu$m we  do not  observe  the strong  drop at  $l\sim120^{\circ}$
(seen at  1.2$\mu$m and in RC stars  at $R_{\odot}\le~2$ kpc) and this
is replaced by a global, and smooth warp signature.
This effect is  further demonstrated when  comparing the warp obtained
for RC stars at $R_{\odot}\le~2$ kpc with that due to the dust (traced
at  $240\mu$m).  The   dust   warp   does  not  show   the   drop   at
$l\sim120^{\circ}$ anymore, and  although still fluctuating, overall a
global and large-scale structure (the warp) is identified.
We emphasize however that a  simple comparison between  the warp of RC
stars  at $R_{\odot}\le~2$ kpc with that  at  $240\mu$m is rather {\em
improper}, since the   DIRBE dust warp  is sensitive   to more distant
regions.
The   comparison has been  meant to   show how  increasing the adopted
wavelength unveils more distant regions and a larger warp amplitude.

Before drawing our   conclusions on the  RC  stars as a stellar   warp
tracer, we analyze the   impact of their  contamination status  on the
global warp properties.
This is illustrated in Fig.~\ref{f_fig4b},  which displays the stellar
warp  as derived   from  RC stars   at heliocentric distances  between
$2\div4, 4\div6, 6\div8$ and $\le8$ kpc.
The upper panels of Fig.~\ref{f_fig4b} prove again how the presence of
a global warp can be debatable for the ``nearest'' RC samples.
Things change considerably for RC stars between $6\le~R_{\odot}\le~8$.
At    this  distance a  global  warp   signature   is  evident in both
hemispheres.
The warp is regular  and shows a  strong asymmetry in the maximum warp
amplitude.   At $l\simeq210^{\circ}$  the   warp brings  the  Galactic
mid-plane $\sim2.7^{\circ}$  below the nominal  $b=0^{\circ}$, whereas
in the   Northern hemisphere  the  maximum  warp   amplitude stops  at
$b\simeq+1.5^{\circ}$.
Comparing  the warp obtained  for   the $6\le~R_{\odot}\le~8$ kpc   RC
sample with DIRBE data shows excellent agreement with that obtained at
240$\mu$m tracing the dust.
This indicates that the  RC star sample  between $6\le~R_{\odot}\le~8$
kpc (although it includes  up to 30\%  contamination by  local dwarfs)
still enables sufficient isolation of distant  stars, those whose warp
signature matches that of DIRBE at 240$\mu$m.

The impact of  contamination on the retrieved  warp  amplitude is best
illustrated in the lower right panel of Fig.~\ref{f_fig4b}.
With  respect   to   the  warp  derived     from  RC   stars   between
$6\le~R_{\odot}\le~8$     kpc, those between $0\le~R_{\odot}\le~8$ kpc
show {\em a  significant change in  the amplitude of  the warp maximum
and location}.
In   particular,    the  maximum    warp     amplitude   passes   from
$\sim-2.7^{\circ}$   (for RC   between $6\le~R_{\odot}\le~8$  kpc)  to
$\sim-1.5^{\circ}$ (for RC at $0\le~R_{\odot}\le~8$ kpc).
A comparison with DIRBE  data now shows  a  better agreement with  the
$3.5\mu$m data and this is again the expected result.
Indeed, both the 2MASS  $0\le~R_{\odot}\le~8$ kpc star counts  and the
DIRBE $3.5\mu$m surface photometry are sensitive all along the line of
sight up to $\sim8$ kpc.

In       conclusion,     the      plots          presented          in
Figs.~\ref{f_fig4a},~\ref{f_fig4b} highlight the  fact that a thorough
characterization   of  the Galactic  stellar  warp,  in particular its
maximum amplitude and maximum  location,   sensitively depend on   the
distance and contamination-status of the employed tracer.
The  lower  panels of Fig.~\ref{f_fig4b} are   the most significant in
demonstrating how  the maximum stellar  warp can shift its location by
$\sim40^{\circ}$ in longitude and can halve its amplitude depending on
the contamination degree of the adopted tracer.
In  addition, the lower right panel   of Fig.~\ref{f_fig4b} proves the
validity  of our  method in tracing  the  {\em stellar  warp}, and the
consistency of the comparison between DIRBE and 2MASS.
This being established we will now focus our discussion on the stellar
warp as    derived    using   red   giants.   As    anticipated     in
Sect.~\ref{s_RC_RGB}, the red giant samples include less contamination
by  local dwarfs and enable a  better and  more reliable separation of
more distant stars.
%

%------------------------------------------one column figure
\begin{figure*}%[h]
\centering\includegraphics[width=14.5cm,height=14.5cm]{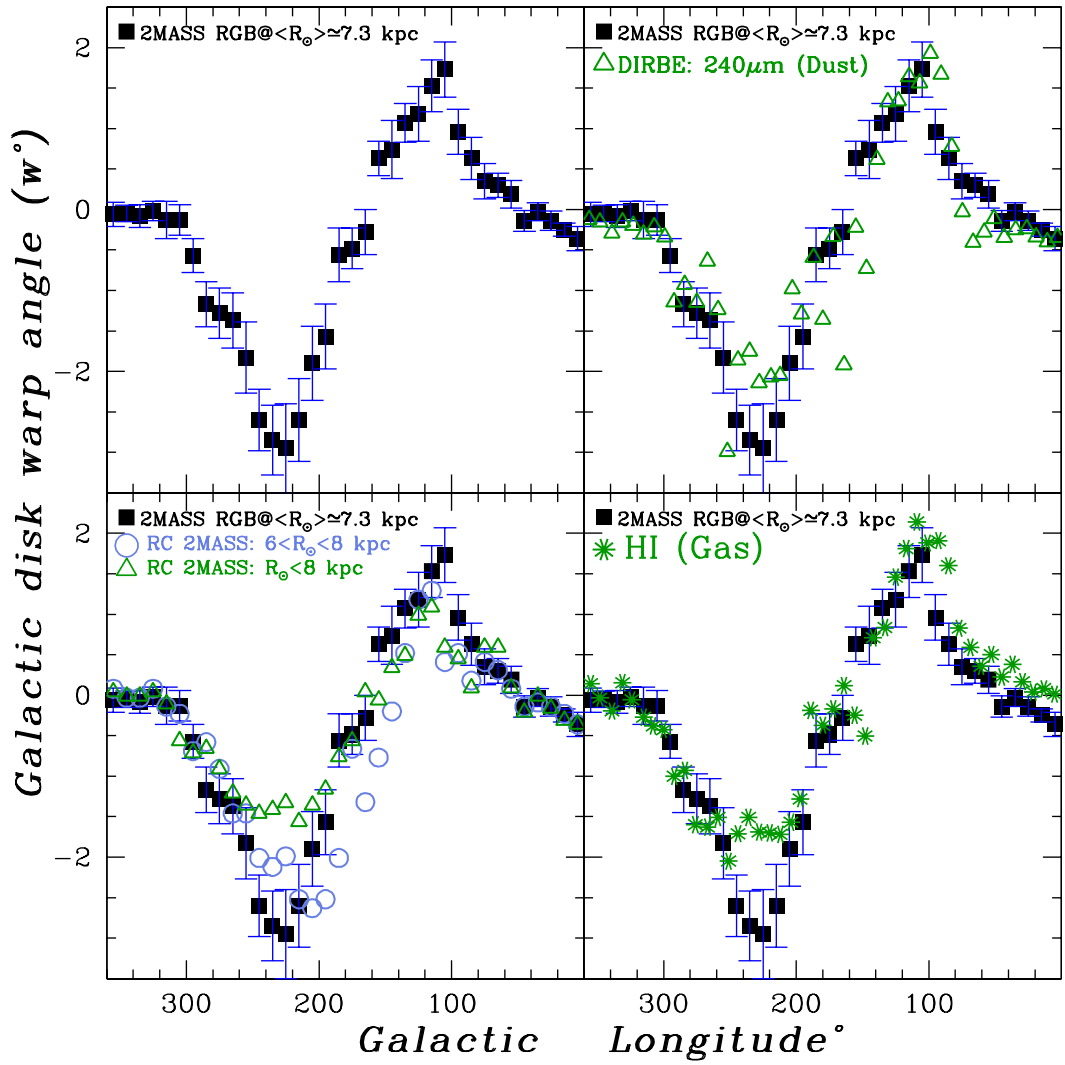}
\centering\includegraphics[width=14.5cm,height=7.5cm]{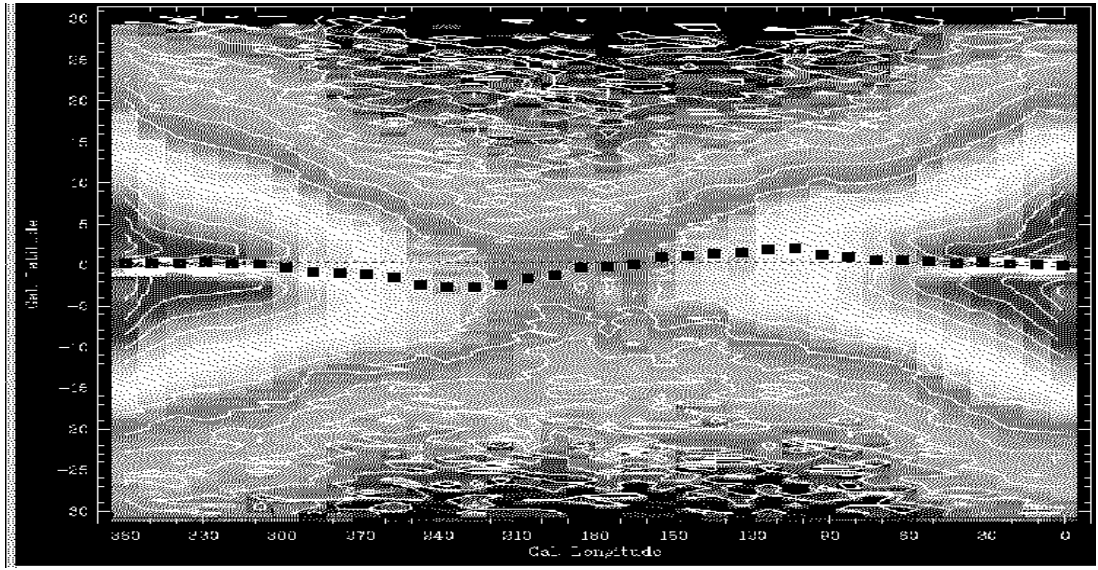}
%----referee
%\centering\includegraphics[width=10.5cm,height=10.5cm]{./psfiles/afig9.ps}
%\centering\includegraphics[width=10.5cm,height=6.5cm]{./psfiles/afig10.ps}
\caption{The upper left panel displays the stellar warp as derived
from red giants at $R_{\odot}\simeq7.3$ kpc. 
In the lower  left panel we compare the  derived warp with those based
on  two RC   samples   between:  (i) $6\le~R_{\odot}\le8$,   and  (ii)
$0\le~R_{\odot}\le8$ kpc.
The upper right  panel over-plots the dust warp  as derived from DIRBE
at $240\mu$m  data,   whereas the lower  right   panel  over-plots the
Galactic warp as derived from neutral \hi\ gas (also from Freudenreich
et al. \cite{freud94}).
The lower panel shows the density maps and contours of the
$R_{\odot}\simeq7.3$ RGB sample.  One can trace by eye the warp
signature (a colored version is more appropriate).  Over-plotted is
also the location of the mean mid-plane warped disk as a function of
longitude.}
\label{f_fig5}
\end{figure*}
%-------------------------------------------------------------

%
\section{The stellar warp as traced by RGB stars}
\label{s_rgb_warp}
\begin{figure}[ht]
\centering\includegraphics[width=8.5cm,height=8.5cm]{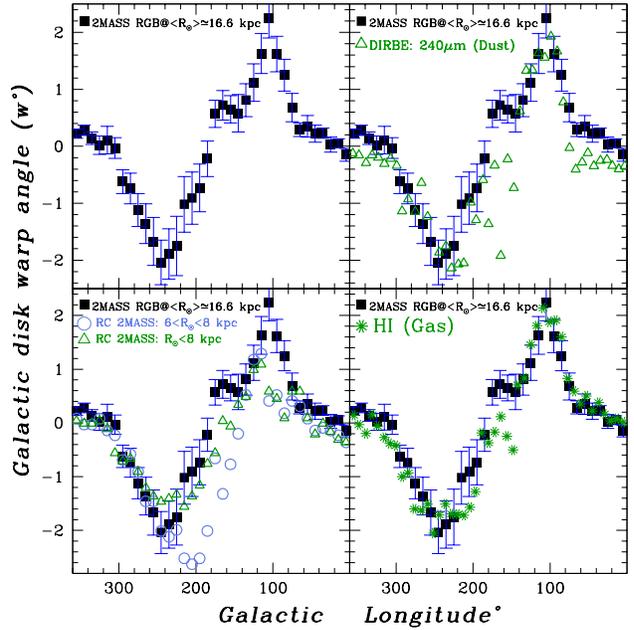}
\caption{Same as Fig~\ref{f_fig5}, but for RGB stars at
$R_{\odot}\simeq16.6$ kpc.  Note  the  structure (``swelling'') around
$l\sim160^{\circ}$ and how the the agreement with  \hi\ is better with
respect to the $R_{\odot}\simeq7.3$ kpc sample.}
\label{f_fig6}
\end{figure}

Figures~\ref{f_fig5} and ~\ref{f_fig6}  display  the  stellar  warp as
derived from RGB samples at $R_{\odot}=7.3$ and $R_{\odot}=16.6$ kpc.
A global regular warp signature is clearly evident in both samples,
reflecting {\em large-scale Galactic structure} (see
Section~\ref{s_w_talk}).
The  most striking feature for  the RGB  $R_{\odot}=7.3$ kpc sample is
the amplitude of the Southern maximum  warp, which lowers the Galactic
mid-plane by almost  $3^{\circ}$.  By applying  a Gaussian fit to  the
data-points   between $160^{\circ}\le~l~\le310^{\circ}$, we estimate a
mean   warp   maximum   at $l\sim235^{\circ}$.   Similarly,   for  the
$R_{\odot}=16.6$ kpc   RGB sample  the warp  maximum  is estimated  at
$l\sim240^{\circ}$.

The  lower left panel  of Fig.~\ref{f_fig5} shows a comparison between
the  stellar     warp   as derived   from     the RC     star  samples
($6\le~R_{\odot}\le8$  and  $0\le~R_{\odot}\le8$  kpc)   with the  RGB
sample    at $R_{\odot}=7.3$ kpc.  This    comparison  is an excellent
demonstration  of how  (i) the  distance   and (ii) the  contamination
status  of the adopted  stellar  tracer affect  the discussion  on the
location and amplitude of the maximum stellar warp.
The  RC  sample between  $6\le~R_{\odot}\le8$  kpc shows  a comparable
Southern warp maximum amplitude with the RGB sample at $R_{\odot}=7.3$
kpc,  and  this reflects the  fact   that  both samples  refer to  the
same distance range.  Yet, the location   of the warp   maximum for the  RC
sample  between $6\le~R_{\odot}\le8$  kpc is $\sim20^{\circ}$  shifted
with respect to that for the RGB sample at $R_{\odot}=7.3$ kpc.
On the other  hand,  the {\em  more  contaminated}  RC sample  between
$0\le~R_{\odot}\le8$ kpc shows a better  agreement with the RGB sample
at $R_{\odot}=7.3$ kpc in the mean location of the warp maximum around
$l\sim240^{\circ}$.   However, the maximum amplitude  of the RC sample
between $0\le~R_{\odot}\le8$ kpc  has clearly decreased, a reminder of
how the  inclusion of nearby  stars  leads to  an under-estimated warp
amplitude.  

We now  turn  our attention to   the  comparison of the  stellar  warp
(derived   from  the RGB  sample at   $R_{\odot}=7.3$  kpc)  with that
obtained  for  the  interstellar   dust and   neutral  atomic hydrogen
components. The upper right panel shows the  comparison with the DIRBE
data at $240\mu$m. The agreement between the stellar  and dust warp is
evident. The two  share the same warp  phase-angle and show only small
differences in the Southern maximum amplitude.
A recent upgrade of the Drimmel \& Spergel (\cite{drimmel01}) Galactic
dust  model, using DIRBE   $240\mu$m  data, shows the  presence  of an
extended spiral arm entering the  third quadrant and intersecting the 
suggested location of CMa.
The reconstruction of  the Galactic dust  distribution is always model
dependent. However, the new model (Drimmel \cite{drimmel05}) indicates
that the far  infrared DIRBE $240\mu$m data,  are sensitive to at least
$\sim7$ kpc from the Sun, and possibly beyond.
The distance consistency of the far infrared DIRBE $240\mu$m data with
the RGB  star sample at $R_{\odot}=7.3$  kpc (and given the excellent
agreement presented in  the upper  right panel of   Fig.~\ref{f_fig5})
allows us  to ascertain that  the dust and  stars at CMa distances are
similarly warped.

Besides the excellent  agreement between the dust  and stellar warp, a
similar conclusion is also obtained for the  gaseous and stellar warp.
The \hi\ latitude of peak brightness as derived by Freudenreich et al.
(\cite{freud94}\footnote{Note  that    although  the Freudenreich   et
al. paper was based on DIRBE data, they also  derived the gaseous warp
based on   a re-analysis of  the  velocity-integrated HI emission maps
from the survey of Weaver \& Williams (\cite{weaver73}).})
follows the same  warp signature we have  obtained from the RGB sample
at CMa distance. The only obvious difference is related to the maximum
Southern warp amplitude of \hi\  that seems to {\it saturate}; showing
an almost constant $w$ warp-angle of $-1.75^{\circ}$ for longitudes between
$210^{\circ}$ and $270^{\circ}$.
Might this  observed difference between the   amplitude of the stellar
and  gaseous warp suggest an ``accretion-perturbation''  on top of the
large-scale warp  structure derived in this paper?.   In regard to the
reality of this effect, one  must keep in mind  that over the last few
years     there  has  been  alternating   evidence   on warp amplitude
differences between stars and gas.
Indeed, we remind the reader that the {\em disk radial truncation}
hypothesis at $R_{GC}\simeq14$ kpc was first proposed (Freudenreich et
al. \cite{freud94}, and Porcel, Battaner \& Jimenez-Vicente
\cite{porcel97}) in order to explain {\em ``why the stellar warp
seemed half the amplitude of  the gaseous warp''}.  Thus, our  results
which   go   in  the   opposite    direction  (stellar RGB    warp  at
$R_{\odot}\simeq7.3$ kpc showing twice the \hi\ warp amplitude) should
only be read in terms of the different confusion status, and therefore
of the mean probed distances, of the gas.

%\centering\includegraphics[width=8.5cm,height=8.5cm]{./psfiles/afig11.ps}
%\centering\includegraphics[width=12cm,height=5cm]{./psfiles/mappa_CM_down.ps}
%------------------------------------------one column figure
%-------------------------------------------------------------
%-------------------------------------------------------------

Indeed, whereas we are able  to isolate the  warp signature  due only
to the stellar  populations  at $R_{\odot}\simeq7.3$ kpc,  the gaseous
warp derived by  Freudenreich et al.   (\cite{freud94}) is that due to
the  summing of different {\it  warp   signatures} by gas  distributed
along the line of sight, at different distances.
This is a   direct   consequence of the   inability to    recover  gas
distribution distances in   lines of sight without   clear kinematical
signatures.
Thus the {\it  saturation}  effect seen in the  \hi\  warp may reflect
that  seen for    the  stellar  warp  (cf    the  lower  panels   in
Fig.~\ref{f_fig4b} and the lower left panel in Fig.~\ref{f_fig5}) when
including nearby stars in distant star samples.
More fundamentally, leaving apart the {\it flattening} of the gas warp
in the range $210^{\circ}\le~l~\le250^{\circ}$, one notes excellent
agreement between of the gas and stellar warp profiles {\em at all}
the other viewing angles.

%----
Moving     to    the stellar  warp     derived     by  RGB  stars   at
$R_{\odot}\simeq16.6$  kpc  (Fig.~\ref{f_zA})  one  notes    structure
(``swelling'') between  $140^{\circ}\le~l~\le180^{\circ}$.   Is this a
deviation from a smooth global  warp signature?.  We suggest that this
is  the same effect seen for  nearby RC samples,  and it is due to the
Orion-Cygnus Local   Arm (short  dashed  line in  Fig.~\ref{f_rings}).
This ``swelling'' is  not seen in  the $R_{\odot}\simeq7.3$ kpc sample
because the selected RGB stars are bright enough to avoid the faintest
regions of the diagrams (mostly populated by nearby dwarfs).
On the  contrary, the RGB sample at  $R_{\odot}\simeq16.6$  kpc can be
easily contaminated by nearby $R_{\odot}\le2$ kpc dwarf stars [compare
the upper and lower panels of Fig.~\ref{f_fig2}] which would enter our
selection box if subject to high reddening  or photometric error.
In turn, the structure appears in the $R_{\odot}\simeq2.8$ kpc RGB
sample (Fig.~\ref{f_zA}) supporting our interpretation that the
``swelling'' is due to nearby dwarf stars that contaminate
primarily the $R_{\odot}\simeq16.6$ kpc RGB sample.
Given the similarities between the $R_{\odot}\simeq2.8$ and $16.6$ kpc
warps,  to  what  extent   might the  $R_{\odot}\simeq16.6$   kpc warp
signature be  in  fact a signature   of  nearby contamination  in  the
$R_{\odot}\simeq16.6$ kpc RGB sample?  The amplitude difference in the
Northern warp  of  the  $R_{\odot}\simeq2.8$ and  $16.6$  kpc  samples
however argues against  this possibility (compare Figs.~\ref{f_zA} and
~\ref{f_zB}).

Recently McClure-Griffiths et al. (\cite{mcclure04}) presented an \hi\
study from  the Southern Galactic Plane  Survey (SGPS), and pointed to
the possible presence of  a distant spiral arm  in the fourth quadrant
of the Milky Way.   The distinct and cohesive  feature  has been traced  over
$70^{\circ}$   and is located    between $18\le~R_{GC}\le24$ kpc.  
In regards to this, it is interesting to  note that our detection of a
stellar   warp in the  RGB  sample at  $R_{\odot}\simeq16.6$ kpc might
represent the  stellar counter-part of  this distant \hi\ spiral arm.
Indeed, the RGB sample at a mean distance of $R_{\odot}\simeq16.6$ kpc
probes the stellar populations between $13.2\div20.0$ kpc from the Sun
(Fig.~\ref{f_rings}) and this is compatible with  the mean position of
this  \hi\ arm  being  within $\sim20$ kpc   from  the Galactic center
(cf Figure.~3a of McClure-Griffiths et al. \cite{mcclure04}).

Interestingly, McClure-Griffiths et al.  find that this distant arm is
well-confined to  the Galactic plane, dropping at  most by $\sim1$ kpc
below  the  Galactic equator.   This again is    in agreement with our
result   (Fig.~\ref{f_zB})  that  the  Galactic   mid-plane (at  these
distances and   lines of  sight) is  located  at $\sim0.5$  kpc  below
latitudes of $b=0^{\circ}$.
The SGPS \hi\ survey analysis however remains limited to
$l\sim250^{\circ}$ (McClure-Griffiths et al.  \cite{mcclure05}), so
that a detailed comparison of the stellar-gaseous warp and the
interpretation of this part of the outer disk awaits the completion of
the third quadrant.
In conclusion,  and as we   shall argue in Section~\ref{s_flare},  the
detection of the warp  in the $R_{\odot}\simeq16.6$ kpc sample  proves
that the MW stellar disk is not truncated at $R_{GC}\simeq14$ kpc, and
that a more extended stellar component is present.
%
% TBD: Mcclure et al.  give a  min. disk density of 
%          $6\times10**{-3} M_{\odot} pc**{-3}$
%          to confine the HI feature scale-height.

%------------------------------------------one column figure
\begin{figure*}[ht]
%\centering \includegraphics[width=8.5cm,height=8.5cm]{./psfiles/afig13.ps}
\centering \includegraphics[width=14.5cm,height=14.5cm]{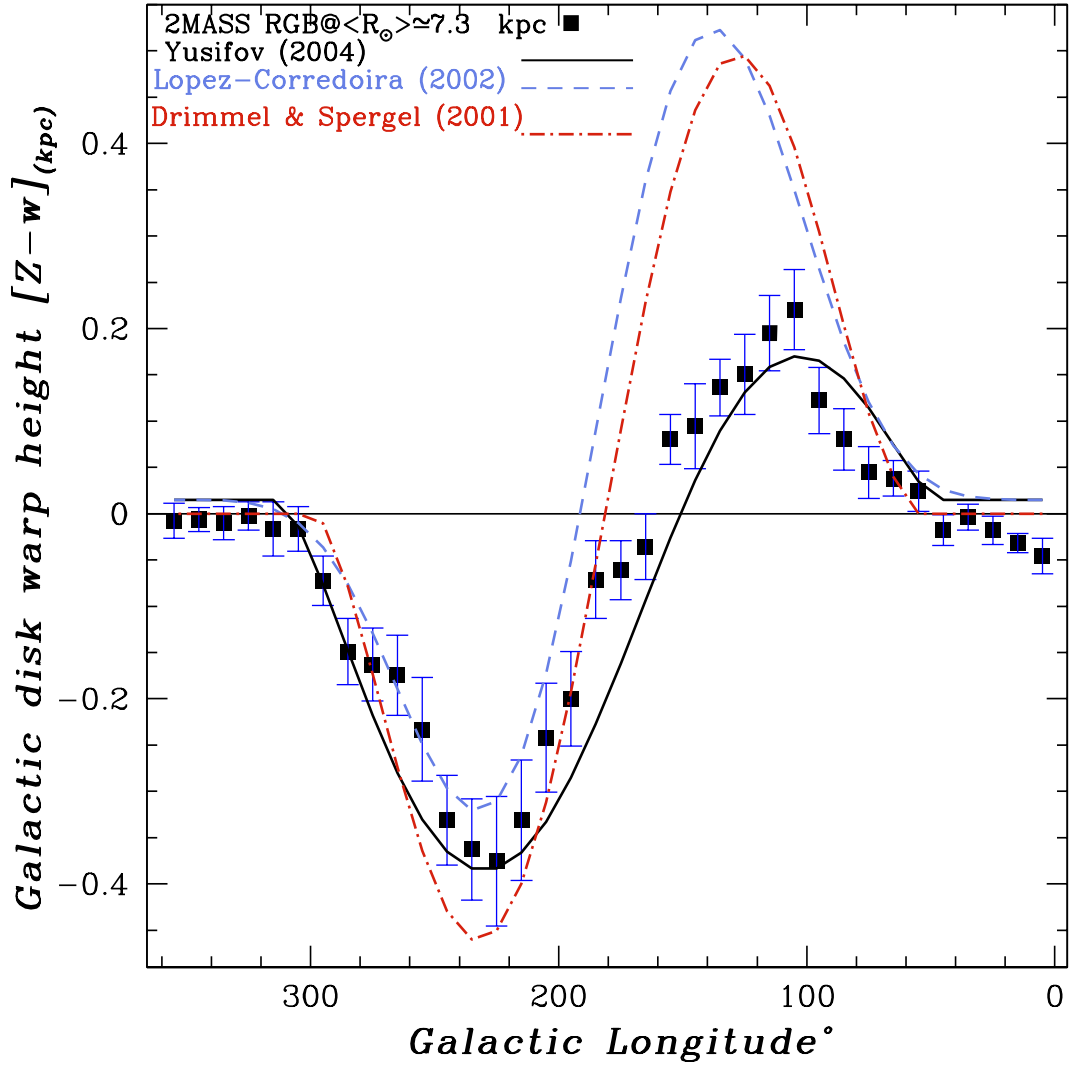}
\caption{A comparison of our  Galactic stellar disk warp height
$Z-warp$ (kpc), as obtained from the RGB sample at heliocentric
distances of $7.3$ kpc points with error bars, with models based
on stellar tracers lines.  Note the excellent agreement between
data and models for the Southern warp. The disagreement for the
Northern warp is explained in the text.}
\label{f_fig7}
\end{figure*}
%-------------------------------------------------------------

\subsection{Comparison with warp models}

Before discussing the possible  implications of the warp as determined
empirically here on the CMa over-density, we first compare our results
with other available models.
In Fig.~\ref{f_fig7} we show a comparison of our derived $Z-warp$ at
the proposed distance of CMa ($R_{\odot}\simeq7.3$ kpc) with
available models based on stellar tracers, namely Drimmel \& Spergel
(\cite{drimmel01})\footnote{Note that in this paper we do not use the
Drimmel \& Spergel (\cite{drimmel01}) model that is based on dust,
since this predicts very high warping (cf  Fig.  2 of Yusifov
\cite{yusifov04}).}, L{\'o}pez-Corredoira et al.  (\cite{lopez02}) and
Yusifov (\cite{yusifov04}).
The 3 models used the distribution of integrated star light, red clump
and pulsar  stars, respectively.  In  the  following all  models  were
converted to $R_{\odot}=8.5$ kpc.
At first sight, one  sees how all  three models converge on showing  a
warp maximum near $l=240^{\circ}$ rather than $l=270^{\circ}$, a clear
hint of the association of CMa with the warp  maximum amplitude.  Most
importantly, at a heliocentric distance of $7.3$ kpc the 3 models show
excellent  agreement     with   our    observed    warp
amplitude around $l\sim240^{\circ}$.

One also notes that there exists: (i) an overall good agreement
between the L{\' o}pez-Corredoira et al.  (\cite{lopez02}) and
Drimmel \& Spergel (\cite{drimmel01}) models; (ii) a significant
dis-agreement between these two models and the Yusifov
(\cite{yusifov04}) model; and (iii) significant agreement between
our observed stellar warp and the Yusifov (\cite{yusifov04}) model.
In Sect.~\ref{s_yusifov} we showed that the Yusifov (\cite{yusifov04})
model is  based  on a  rather  incomplete  catalog,  thus the   closer
agreement  of our retrieved warp   with this model  was an un-expected
result.  To solve this confusion however, one must go into the details
of the models.

The 3 models describe the warp  as a series  of concentric rings which
intersect the  Galactic plane along  a line of nodes (usually measured
as  the Galactocentric  angle from   the  sun to the Galactic   center
$\phi$,   see  Fig.~\ref{f_rings}    and     Fig.~1A of   Evans     et
al. \cite{evans98}).
The   models of  L{\' o}pez-Corredoira  et   al.  (\cite{lopez02}) and
Drimmel   \& Spergel  (\cite{drimmel01})  report $\phi=-5^{\circ}$ and
$0^{\circ}$   respectively.    On the     other  hand,  the    Yusifov
(\cite{yusifov04}) model derives a rather higher $\phi=+15^{\circ}$.
$\phi$ different than zero means that  the Sun is not located on
the line of nodes, instead it is already {\it  inside} one of the
two oppositely warped regions.
As  a  consequence, when viewed from  the  Sun, geometrically the warp
maximum  is {\it nearer}  to us in  one  hemisphere than  it is in the
other.
Thus, if the   warp is traced  at  fixed distances  from the  Sun  and
$\phi\neq0$, then the {\em observed} warp is {\em asymmetric}.
Yusifov (\cite{yusifov04})  derives a {\em  positive} $\phi$ value and
therefore predicts a significantly   asymmetric warp, with a  relatively
stronger Southern  warp maximum   amplitude.
On the contrary, L{\' o}pez-Corredoira et al.  (\cite{lopez02}) derive
a {\em negative} $\phi$  value which produces an observable asymmetric
warp,  with a relatively stronger  Northern warp  maximum amplitude.
Lastly,  Drimmel \& Spergel (\cite{drimmel01}) derive $\phi=0^{\circ}$
and therefore   their model  predicts  a  symmetric  warp;  equivalent
Northern/Southern warp maximum amplitudes.  

Another  factor that  contributes to the  appearance  of an asymmetric
warp is due to  the chance location of  the Northern warp just  behind
the Norma-Cygnus arm (labeled {\em outer} in Fig.~\ref{f_rings}).
Thus,  extinction in the Norma-Cygnus arm  and  possible variations of
the  extinction  curve (due  to the  penetration  of  gaseous or dusty
regions), coupled with the possibility that the Sun may not lie on the
line  of  nodes are all factors   that  conspire in  producing  a less
pronounced  apparent  Northern warp at heliocentric distances of
$2.8$ and $7.3$ kpc.

\begin{figure}[h]
\centering \includegraphics[width=8.5cm,height=8.5cm]{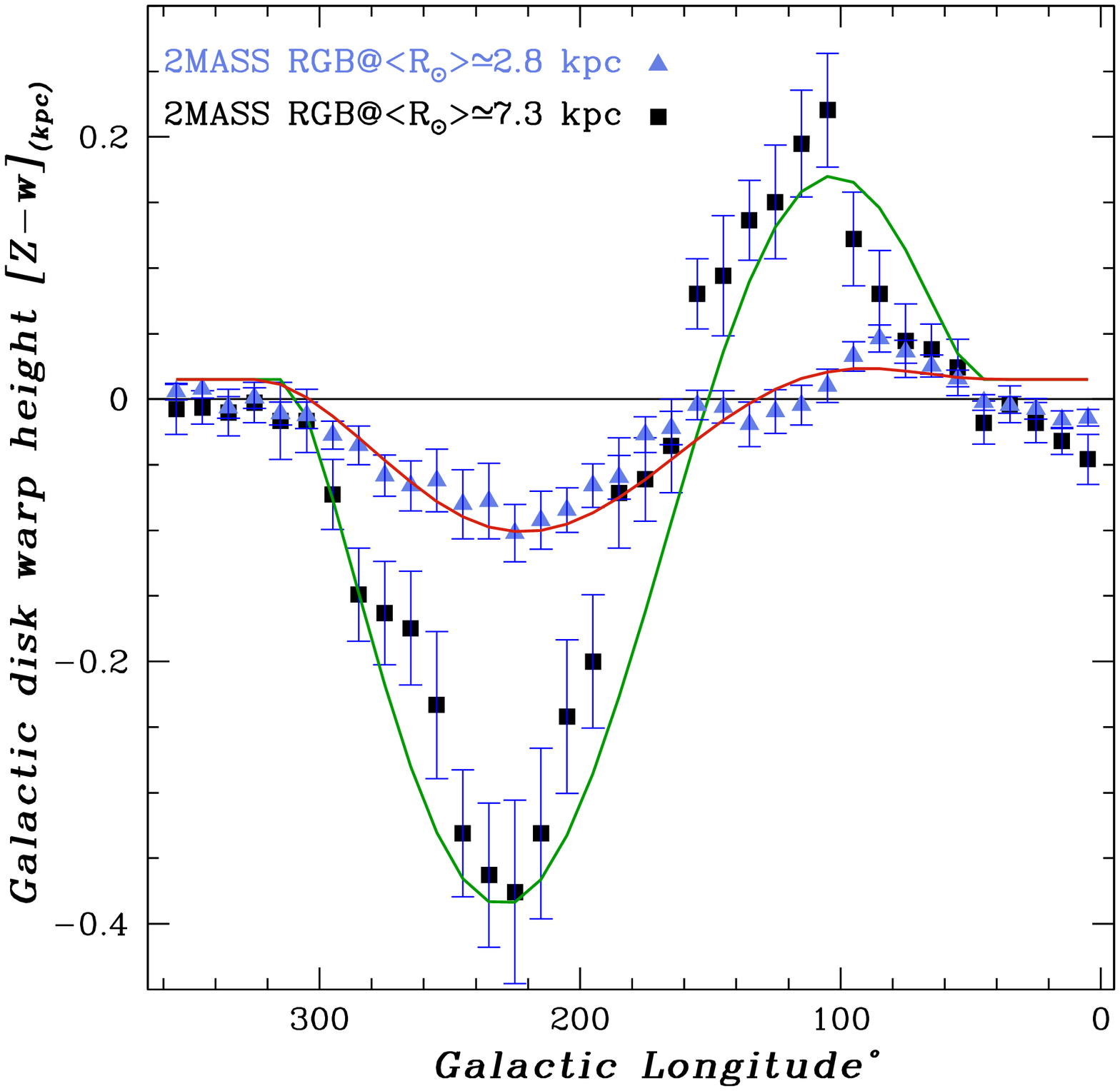}
\caption{The Galactic stellar disk warp height as derived from
the RGB samples at   $R_{\odot}=2.8$   and   $7.3$ kpc   compared    
with the  Yusifov (\cite{yusifov04}) model.}
\label{f_zA}
\end{figure}

To develop further the particular issue  of a symmetric and asymmetric
stellar warp, we now  compare the stellar warp as  traced by the 3 RGB
samples   at   different   heliocentric distances.     Since our
results   favor rather high and positive $\phi$  values
we will continue  the comparison with the  Yusifov  model as this  has
shown the  best fit to our  RGB sample at  heliocentric distance $7.3$
kpc.
Figure~\ref{f_zA} shows  clearly  that the  Yusifov  model
provides a  satisfactory match with  the stellar warp as 
derived by the $R_{\odot}\sim2.8$ and $\sim7.3$ kpc RGB samples.
Figure~\ref{f_zA} also shows how the Northern warp amplitude is almost
half that of the Southern warp, for  the two distances.

\begin{figure}[h]
\centering \includegraphics[width=8.5cm,height=8.5cm]{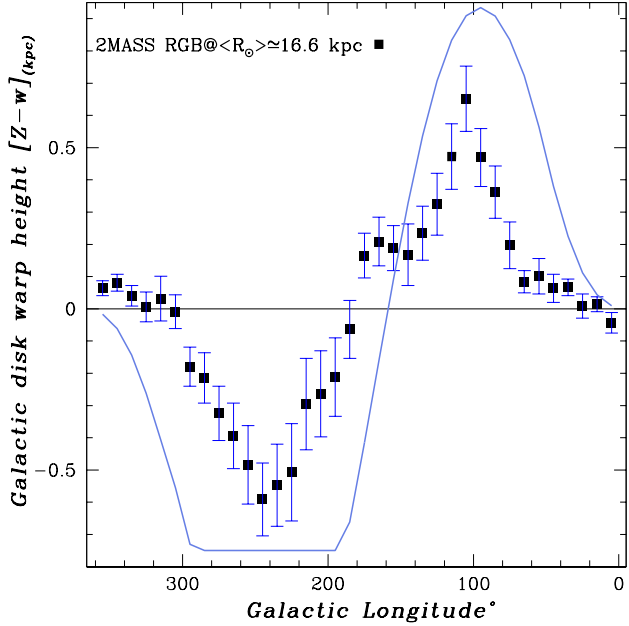}
\caption{The Galactic stellar disk warp height as derived from
the RGB sample at $R_{\odot}=16.6$  kpc {\it improperly} compared with
the  Yusifov   (\cite{yusifov04})  model (not    applicable  at  these
distances).  Note that for Galactocentric   distances greater than  14
kpc the model imposes  a constant Southern  stellar  warp so as  to be
consistent with the derived parameters of the gaseous warp.}
\label{f_zB}
\end{figure}

In principle,  the warp derived at  $R_{\odot}\sim16.6$ kpc should not
be compared with the  Yusifov model (Fig.~\ref{f_zB}) since that  model
is not applicable   at $R_{GC}\ge14$ kpc.   Nevertheless,  for the
derived warp at $R_{\odot}\sim16.6$ kpc we find that the Yusifov model
predicts  maximum warp  amplitudes  that are   not very  far  from our
derived values; with amplitude  differences  of $0.25$ and $0.15$  kpc
above  and below the mid-plane,  respectively.  Moreover, the location
of the   warp  maxima  are still   close  to  those of  the   derived
data-points, indicating that the warp phase angle is still in
agreement with observations.
Figure~\ref{f_zB},  also shows    another important  finding:  only at
distances of   $R_{\odot}\simeq16.6$ kpc can  we   obtain a good
symmetry between Northern  and Southern warp maximum amplitudes;
both being around $\sim0.6$ kpc above and below the mid-plane.
The  fact that we  obtain a symmetric  warp at $R_{\odot}\sim16.6$ kpc
also indicates   that probing the  warp  at  such large  distances  is
affected less by the dependence on $\phi$.

The passage from the stellar warp at $R_{\odot}\sim7.3$ kpc to that at
$16.6$  kpc shows an out-break in  the Northern warp maximum by almost
$0.45$ kpc.  This confirms how probing distant regions (which are less
affected by grand-design structures like the Norma-Cygnus arm) reveals
the real entity of the Northern warp amplitude.
On   the other    hand,  the   passage  to   the    stellar  warp   at
$R_{\odot}\sim16.6$ kpc shows a {\em limited} increase in the Southern
warp maximum amplitude by $\sim0.2$ kpc.
This  {\em limited} increase  probably  reflects the  way the Southern
gaseous  warp shows a {\it saturation}  effect at $\sim0.75$ kpc below
the  plane,   that induced   previous investigations  to   consider it
constant after $R_{GC}=14.0$ kpc (Burton 1998).

%------------------------------------------one column figure
%-------------------------------------------------------------

%------------------------------------------one column figure
%-------------------------------------------------------------

\subsection{Significance of the retrieved stellar warp}
\label{s_w_talk}

Figures~\ref{f_fig5},   ~\ref{f_fig6}, ~\ref{f_fig7}, ~\ref{f_zA}  and
~\ref{f_zB} allow us to conclude the following:

{\em -- The correspondance  of the CMa  over-density with the Southern 
warp maximum at $l\sim240^{\circ}$}:
This conclusion is  based on both our  derivation of the  stellar warp
(at   different  distances)  and  independently  derived  models. This
establishes  that the  warp maximum  is   significantly displaced from
$l\simeq270^{\circ}$ at CMa distances.

Now  that   our   results    indicate  that  the   Galactic    warp at
$R_{\odot}=7.3$ kpc can bring  the mid-plane $\sim3^{\circ}$ below the
nominal $b=0^{\circ}$, star-count   comparisons  above and below   the
plane should be  made   as follows:  in   the Canis Major    direction
($l=240^{\circ}$) star counts at CMa core ($b\simeq-8^{\circ}$) should
be compared with those at $b\simeq+2^{\circ}$.
This explains    the consistent detection   of  a CMa  over-density or
peculiar    signature  when     comparison  fields  are     taken   at
$b\simeq+8^{\circ}$ (Martin et  al.  \cite{martin05} and Bellazzini et
al. \cite{luna04}).
One further piece  of evidence which indicates  that the CMa detection
is  the recovery of the Galactic   warp Southern hemisphere maximum is
found in the so-called {\em structure A}.
Along  with the CMa  over-density, {\em  structure   A} is the  second
over-density  found   in the   Martin    et  al.    (\cite{martin04a})
analysis. Its location   (in the Northern  hemisphere) makes  it fully
compatible with being the Northern warp maximum, as also found in this
paper.

Yet     another overdensity (in Argo)    has   recently been announced
(Rocha-Pinto et  al.  \cite{rocha05}). Our  results  suggest that this
over-density cannot  be {  directly  correlated} to the Galactic  warp
maximum (as suggested in Bellazzini et al. \cite{luna05}).
Indeed, at  $l\simeq290^{\circ}$ ($\sim$Argo center) the Galactic warp
amplitude is less pronounced than that at $l\simeq240^{\circ}$.
To  consider a possible origin  of this over-density,   we
briefly reconsider  the analysis of Rocha-Pinto et al.  (\cite{rocha05}).
In order to increase their accessibility to low latitude sky areas and
reduce  the dependency on  foreground   reddening Rocha-Pinto et   al.
(\cite{rocha05})  use a  Galactic   model  to remove the    foreground
density.  The  model is idealised  in that  it assumes a cylindrically
symmetric   density  distribution   about    the  Galactic     center.
Figure~\ref{f_fig5}  shows that   the   Galactic  warp   amplitude  at
$l\simeq290^{\circ}$ is not zero, inconsistent with the assumptions of
the Rocha-Pinto Galactic model.
Although this effect remains  to be quantified,  and although the warp
amplitude is not extreme at  $l\sim290^{\circ}$, the high star density
this close  to the Galactic   disk will amplify the residual  Galactic
component when Galactic populations are removed, necessarily affecting
the quantification of the Argo over-density.

{\em  -- The  regularity  of    the  Galactic   stellar  warp   out    to
$R_{GC}\sim20$ kpc:}
We have measured the stellar warp  at 3 different distances and argued
that   the  only    visible  perturbation away    from  regularity  (a
``swelling''  in  the  $R_{\odot}=2.8$  and 16.6   kpc rings)   can be
explained as being   due to the  presence of  the Orion-Cygnus  arm in
lines     of   sight    towards   the Northern     warp   and  between
$140^{\circ}\le~l~\le~180^{\circ}$.
That is, we conclude that a {\em  global and regular warp signature is
traced  to Galactocentric distances of  at least $\sim20$  kpc} in the
anti-center direction.  We emphasise this finding, since it is a clear
conclusion  even though there  are two factors  that could have masked
such a signature: 1) the Galactic flare (see next  Section) and 2) the
decreasing stellar density in the Galaxy's outskirts.

{\em  -- The  consistency of the stellar  warp  with the dust and gaseous
warp:}
In addition to being a regular large-scale structure, the stellar warp
is consistent  with that of the interstellar  dust and  neutral atomic
hydrogen.
The consistency might have    been expected given the   close physical
correlation  between   these  3   components.   However,  it   is when
considering  accretion scenarios   that  this  finding   acquires more
significance.
Accretion  of  companions    can  be  responsible  for  generating   a
short-lived warp, and  can  modify differentially  the  stars-dust-gas
warp properties [cf Binney (1992)].
In regards  to this,  it is  interesting to recall  that the numerical
simulations by Helmi et  al.  (\cite{helmi03}), cited to explain
the almost circular orbits  of CMa  and Mon.   Ring, suggest  that the
accretion event must be  relatively  young ($\le1$ Gyr)  otherwise any
coherent  structure   would   be  dissolved.  
Developing consistency between dynamically young events -- a few
rotation periods -- and global regularity remains to be investigated in
detail. 

{\em -- The orientation of the Galactic bar and warp:}
The results presented in this paper define a stellar warp 
with a rather high and  positive phase-angle.
Interestingly, we note  that a positive  warp phase-angle  follows the
orientation   of   the  Galactic  bar.     The  amplitude of  the warp
phase-angle ($+15^{\circ}$)  in  Yusifov's  model is very   similar to
recent determinations  of  the Galactic bar  orientation: Freudenreich
(\cite{freud98}) derive  $+14^{\circ}$,   while Babusiaux  \&  Gilmore
(\cite{babu05}) derive $+22^{\circ}\pm5.5^{\circ}$.
We have  shown that the   pulsar  catalog used  by  Yusifov is  biased
towards the  Galactic  inner    regions  ($R_{GC}\le  8$  kpc,     see
Fig.~\ref{f_fig1}). Whereas this could have been  a shortcoming in his
model, it turns out that this {\it inner regions weighted catalog} may
have led to a better determination of an important warp parameter; its
phase-angle at least in the inner Galaxy.
Thus, it may be that the Milky Way  bar and the inner warp-rings share
the  same orientation angle.   This point will be further investigated
in Momany et al.   (in prep.).  For the time  being, we note that  our
determination   of  a   non-zero phase    angle frees    us  from  the
uncomfortable   assumption of a {\em  fortuitous}  location of the Sun
along the line of nodes.

\section{The Galactic flare}
\label{s_flare}

In the present section, and before extending our  discussion on
the Mon.  Ring,  we derive another  necessary Galactic ingredient, 
flaring of the Galactic disk.

For this  analysis we use  the $ Z_h$  values  of  the  scale height as
determined in   Sec.~2.3    for  the  3    M-giants   samples  at
$R_{\odot}=2.8$, $7.3$ and $16.6$ kpc.
The flaring  of  the Galactic   disk (viewed  as  an increase  of  the
scale-height with  increasing  Galactocentric distance)  is evident in
Figure~\ref{f_F2}, where we trace the radial trend of $Z_h$.
Before discussing   the details  of Figure~\ref{f_F2}, we   remind the
reader that: (i) we used a single  exponential in fitting the vertical
density profile, defining a scale-height;  (ii) in the fitting process
we  excluded regions at  low latitudes so  as to avoid regions of high
reddening    corrections  around the   Galactic   mid-plane;  and most
importantly (iii)   our  three 2MASS  M-giants samples  have  an upper
vertical limit in latitude of  $b=+/-20^{\circ}$, and this corresponds
to different upper $Z$ probing once the  distances of the 3 samples is
accounted for.
This  also means that  we have a  different  weighting of the thin and
thick disk stellar   populations for the   three M-giant  samples.
Table~\ref{t_flare} reports the adopted  limits in the vertical density
profile fitting, for the three samples.
Considering  that at the  solar distance the  scale-height of the thin
disk is $\sim0.3$ kpc while that of the thick disk is $\sim1$ kpc (Wyse
\&   Gilmore \cite{wyse05}),  it    is   clear that  the sample     at
$R_{\odot}\simeq2.8$ kpc is weighted more by thin disk populations. On
the contrary the $R_{\odot}\simeq7.3$ and $16.6$ kpc are weighted more
by thick disk populations.

Thanks to the  use of different lines of  sight at fixed  heliocentric
distances,  our data-points cover    a  wide range of   Galactocentric
distances and  thus are  reliable  in tracing the  flare  all over the
Galactic stellar disk (inner and outer).
In particular we note the following:

%------------------------------------------one column figure
\begin{figure*}[ht]
%\centering
%\includegraphics[width=7.5cm,height=7.5cm]{./psfiles/afig16.ps}
\centering\includegraphics[width=11.5cm,height=11.5cm]{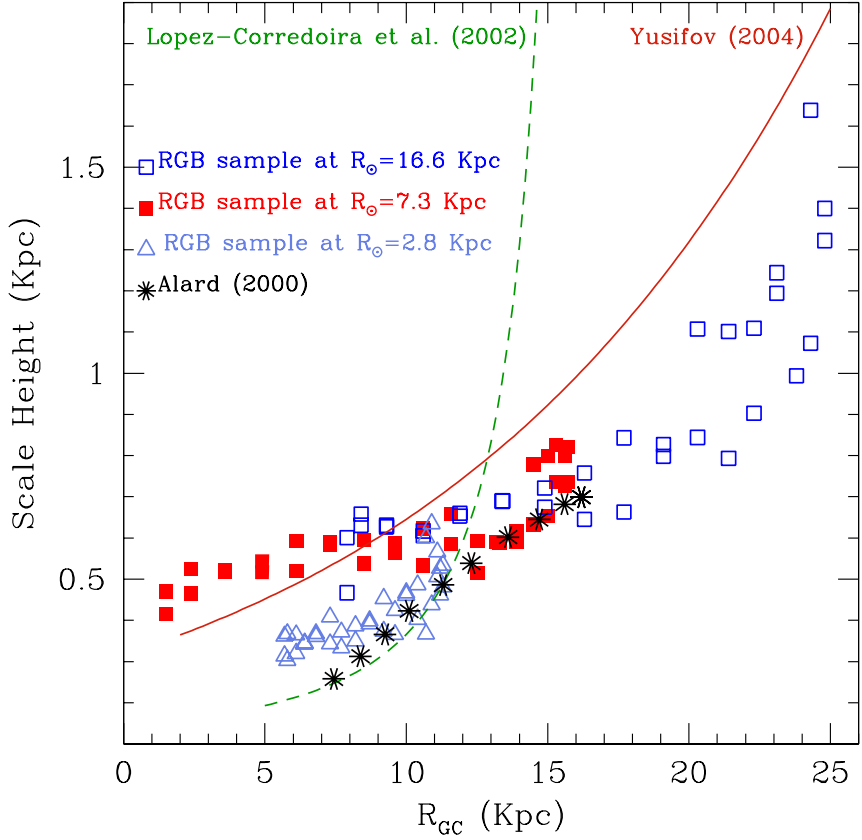}
\caption{Variation of the  scale-height (derived from the
$R_{\odot}=2.8$, $7.3$  and $16.6$ kpc RGB  samples)  as a function of
Galactocentric distances.   We also show  a comparison  of our results
with two models and data-points from Alard (\cite{alard00}).}
\label{f_F2}
\end{figure*}
%-------------------------------------------------------------

\begin{itemize}
\item For Galactocentric distances in the range 
$4\le~R_{GC}~\le15$ kpc, the $R_{\odot}=7.3$  kpc M-giants  sample (filled
squares),  shows only  small  variations  and  in general displays  an
almost constant scale-height.
Interestingly, the last data-points clustering around $R_{GC}\simeq15$
kpc show a saturation of the scale-height at $\sim0.75$ kpc;

\item  For Galactocentric distances in the range 
$8\le~R_{GC}~\le25$  kpc, the   $R_{\odot}=16.6$  kpc M-giants  sample
(open squares), shows a bimodal behavior with an: (i) almost constant
scale-height     within  $R_{GC}~\le15$ kpc;     and   (ii) increasing
scale-height for $R_{GC}~>15$ kpc. 
The inner radii with the $\sim$constant trend show a mean
scale-height of the same order as that derived from the $R_{\odot}=7.3$
kpc M-giants sample.
By contrast,  at  large  Galactocentric distances with   an increasing
scale-height we note that the data-points  do not converge to a single
value,  but show a gradual   increase with distance  out  to at  least
$R_{GC}\sim23$ kpc.
The general trends reflect that seen in Alard (\cite{alard00}, starred
symbols) and show fair agreement with the Yusifov model.
A quantitative parametrization of the observed  flare and warp will be
subject of a future paper (Momany et al. in prep.).

\item  In contrast to the almost constant scale-height derived from the
$R_{\odot}=7.3$ and $16.6$ kpc M-giants samples for the inner regions,
the sample at $R_{\odot}=2.8$ kpc (open triangles) shows some differences.
Between $5\le~R_{GC}~\le9$ kpc, the  $R_{\odot}=2.8$ kpc shows a  mean
scale-height   of $0.35$   kpc,   almost half   that   found for   the
$R_{\odot}=7.3$ and $16.6$ kpc M-giants samples ($0.65$ kpc).

This difference  in  scale-height is  mainly due  to the different
weighting of  the thin and  thick disk populations when extracting the
M-giants       samples       at     different      distances   between
$-20^{\circ}\le~b~\le20^{\circ}$.   This    translates  into different
limits   when   performing       the   scale-height   fitting    (cf
Table~1).  For  the $R_{\odot}=2.8$ kpc sample the profile
fitting was  made between  $0.15$  and $0.85$ kpc, clearly more weighted
towards the thin disk regime. Reassuringly, the overall mean
scale-height of the $R_{\odot}=2.8$ kpc  sample ($\sim0.35$) is  close
to typical values for that of the thin disk.
On the other hand, for the  $R_{\odot}=7.3$ and $16.6$ kpc samples the
profile   fitting    starts    at     $\sim1.5$ times   the  thin      disk
scale-height. Therefore,   the   overall  mean   scale-height   of the
$R_{\odot}=7.3$ and $16.6$   kpc   samples is an intermediate    value
($0.65$ kpc) that reflects the  mixing of the thin  and the thick disk
populations and respective scale-heights.
Lastly, besides the  small difference  in mean scale-height, one  notes that
the sample  at $R_{\odot}=2.8$ kpc shows a  similar trend with
Galactocentric distance as does  the flare
model derived by L{\' o}pez-Corredoira et al.  (\cite{lopez02}).

\end{itemize}

\begin{table}
\begin{center}
\caption{The limits within which we fit the vertical scale-height for
the three M-giants samples. All values are in kpc.}
\begin{tabular}{|r|c|c|}
\hline
Distance & Lower  $Z$ limit & Upper $Z$ limit \\
\hline
  {\bf 2.8}  &   0.15   &   0.85 \\
  {\bf 7.3}  &   0.45   &   2.25 \\
 {\bf 16.6}  &  0.60   &   3.00 \\
\hline
\end{tabular}
\end{center}
\label{t_flare}
\end{table}

In   general, our data-points  reflect the  fact that the scale-height
increases with increasing Galactocentric radii.
For the inner Galaxy the scale-height is rather similar  to the that in
the solar     neighborhood, a consistency  that  continues   to  about
$R_{GC}\sim15$ kpc.
Further   out the scale-height grows   relatively rapidly, with a
smooth  increase until about $\sim23$ kpc.

The thickening of the stellar disk is the re-distribution of stars
from lower to higher latitudes. A disk thickening or flaring is a
phenomenon that requires a reduction below single exponential
extrapolation in the stellar density near the Galactic plane for the
outer disk.
Thus,  our  detection of   (i)  a  large-scale  warp signature  out to
$R_{GC}\sim24$ kpc; and (ii) a  disk flaring out to $R_{GC}\sim23$ kpc
provides new evidence of a rather extended Milky Way stellar disk.
Therefore,   our results  add       to earlier suggestions  by    L{\'
o}pez-Corredoira  et al.  (\cite{lopez02})  and Alard (\cite{alard00})
that  the  thin disk  cut-off  at  $R_{GC}\sim15$  kpc  (Ruphy  et al.
\cite{ruphy96}) {\em ``is  not only unnecessary but  also inconsistent
with the data''}.

\begin{figure}
\centering\includegraphics[width=7.5cm,height=4.5cm]{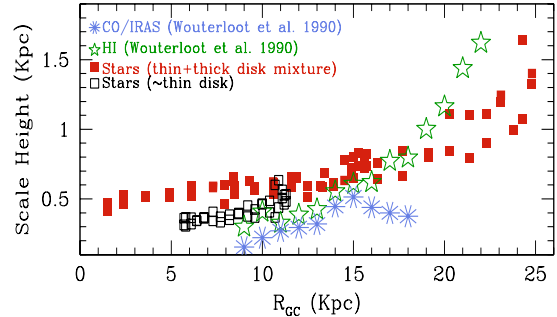}
\caption{A comparison of the thickness of the stellar disk, neutral
hydrogen gas  layer and molecular clouds ensemble  for the outer Milky
Way. Filled squares show the flaring for the RGB samples at
$R_{\odot}=7.3$ and $16.6$ kpc, whereas the open squares plot the RGB  samples at
$R_{\odot}=2.8$ kpc.}
\label{f_flare_HI}
\end{figure}

In  Fig.~\ref{f_flare_HI} we compare   the thickening/flaring  of  the
stellar disk (as obtained from the  RGB samples at $R_{\odot}=7.3$ and
$16.6$  kpc) with that found  for the neutral  hydrogen  gas layer and
molecular clouds ensemble (Wouterloot et al. \cite{wouter90}).
On    the  one     hand,   the  observations   by  Wouterloot    et
al. (\cite{wouter90}) are a reminder  of the presence of  molecular
clouds with  embedded  star formation out  to $R_{GC}\sim20$  kpc. The
authors argue that the  lack of CO emission  at $R_{GC}>20$ kpc is not
to  be  attributed to the sensitivity   of the IRAS survey,  rather it
shows the absence of recent star formation at these distances.
On the other hand, the Figure proves the  compatibility of the stellar
and gaseous  flaring for  the outer  Milky Way.  We  remind the reader
that the filled squares refer to a  mixture of thin/thick disk stellar
populations, and this   explains the  gas-stars differences in relative
scale-heights. The different populations shown have different vertical
velocity dispersions, hence their different scale heights. 
The only unexpected aspect of this comparison is the very high outer
gas scale height, apparently exceeding that of the stars beyond
$R_{GC}\sim20$  kpc. This is not understood.
%

%------------------------------------------one column figure
%-------------------------------------------------------------

\section{The warped and flared Galactic disk and the connection with 
stellar over-densities}

Having demonstrated the existence and determined the properties of the
stellar   warp and  flare, we now   apply  this description  of a {\em
deformed}  stellar disk to  investigate its impact on reported stellar
over-densities at low Galactic latitudes.
In the previous sections we showed that the warp  and flare are fairly
well  described by the  Yusifov model.  We therefore will continue
using this model. We remind the reader that  the model describes rather
well   the  inner $R_{GC}\sim14$ kpc, but represents only a
qualitative description  of   the outer  Milky Way  warped  and flared
regions.
%%%%%%
\subsection{The Canis Major over-density}
\label{s_cma}

\begin{figure}[h]
\centering \includegraphics[width=8.5cm,height=5.5cm]{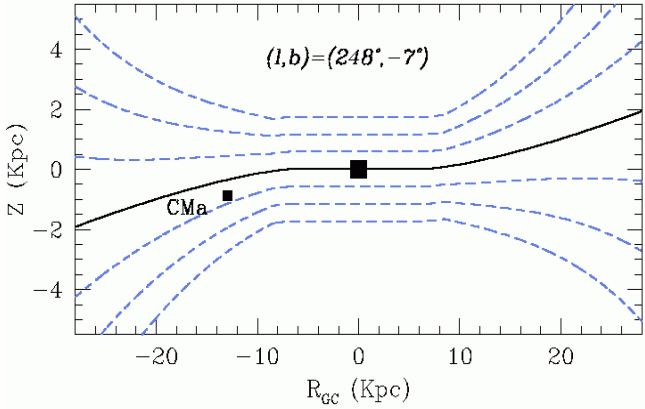}
\caption{A cut in the YZ plane of the warped and flared Galaxy using the
Yusifov warp model in  the direction and at the  distance of CMa.  The
thick line marks the mean warped stellar disk  whereas the grey dashed
lines trace  the density at  1x,  2x and  3x  the scale-height  of the
disk.}
\label{f_F3}
\end{figure}

To show tangibly the importance  of allowing for deviations away  from
symmetry  around $b=0^{\circ}$ for  studies  of the Milky Way  stellar
disk, in Fig.~\ref{f_F3} we  show a cut in  the YZ plane of the warped
and flared  Galaxy\footnote{See also Fig.~1A   and 1B of Evans  et al.
(\cite{evans98}).}  using  the Yusifov  warp  model  in the  direction
[($l,b$)$=248^{\circ},-7^{\circ}$] and  distance [$R_{\odot}\simeq7.2$
kpc]   of  CMa  (as  most     recently  derived  by  Bellazzini     et
al. \cite{luna05}).

The solid  line traces the mean  warped stellar disk, whereas the grey
dashed lines follow the three scale heights of the Galactic disk.  The
Figure   shows clearly that the  CMa  over-density  falls at less than
$\sim500$ pc from the  Galactic  warped mid-plane, that  is $\sim50\%$
the maximum density of the stellar disk at that distance.
%
%3 altezza:  : 5%
%2 altezza:  : 14%
%1 altezza:  : 37%

This figure  clarifies  how  important is  the   Galactic warp to  any
analysis of the  CMa over-density. Specifically, comparing star counts
at $Z\simeq-1$ kpc with those at $Z\simeq+1$ kpc one ends up comparing
a region falling  within $\sim50\%$ of the  stellar  disk density with
another having only $\sim14\%$.
%
%------------------------------------------one column figure
%-------------------------------------------------------------

\subsection{The Monocerous Ring over-densities}
\label{s_mon}

The knowledge of the outer  structure of the  Galactic disk derived in
this  analysis  does  not provide   an  explanation for the Monocerous
Ring. However,  we show that the Galactic  flare {\em can} account for
some aspects of some positive detections  of the Mon.  Ring in various
surveys   and therefore,  needs    to   be  taken  into account     in
investigations  of the relative  importances of  accretion or Galactic
structure for the origin for the Mon. Ring.
%

%------------------------------------------one column figure
\begin{figure}[h]
\centering \includegraphics[width=8.5cm,height=8.5cm]{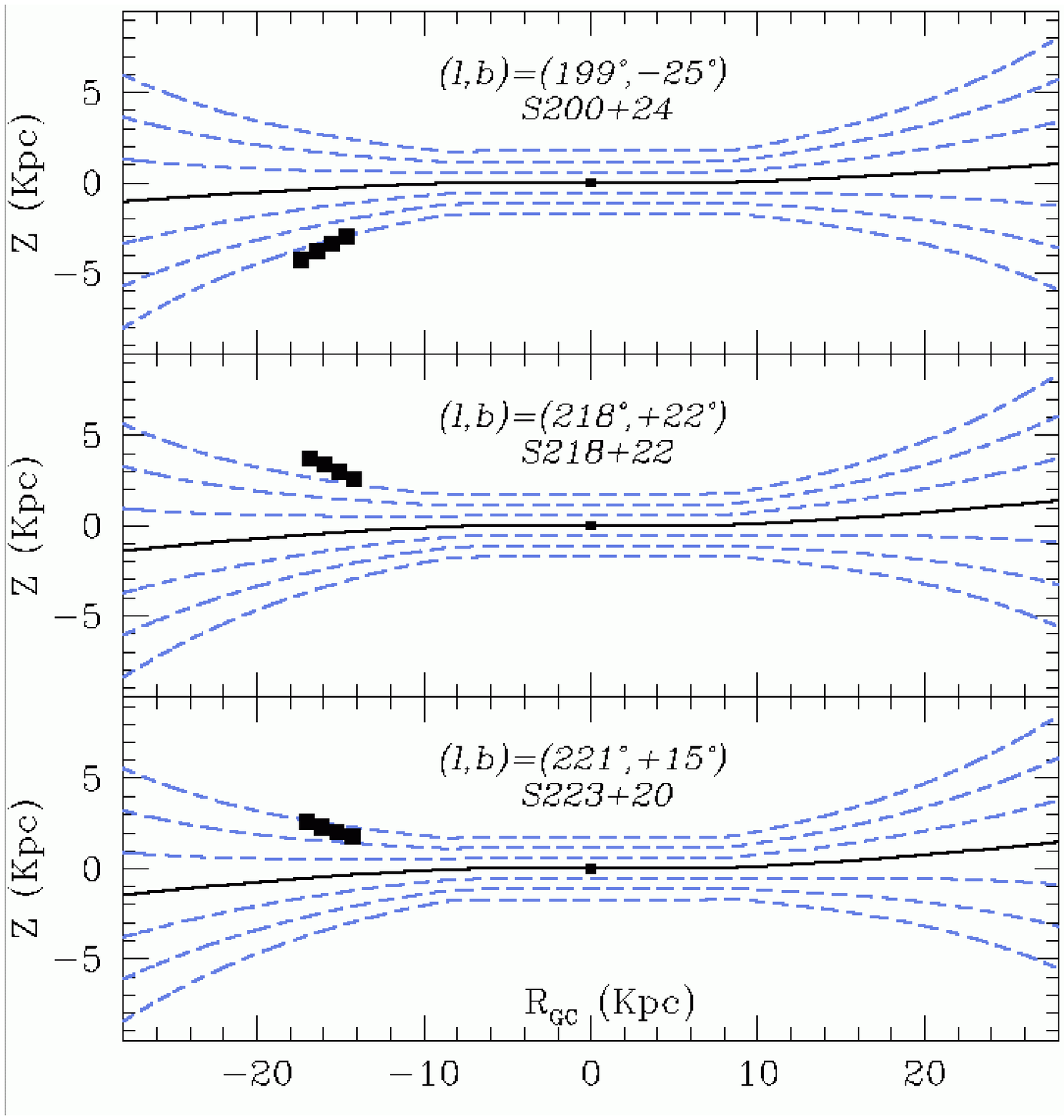}
\centering \includegraphics[width=8.5cm,height=8.5cm]{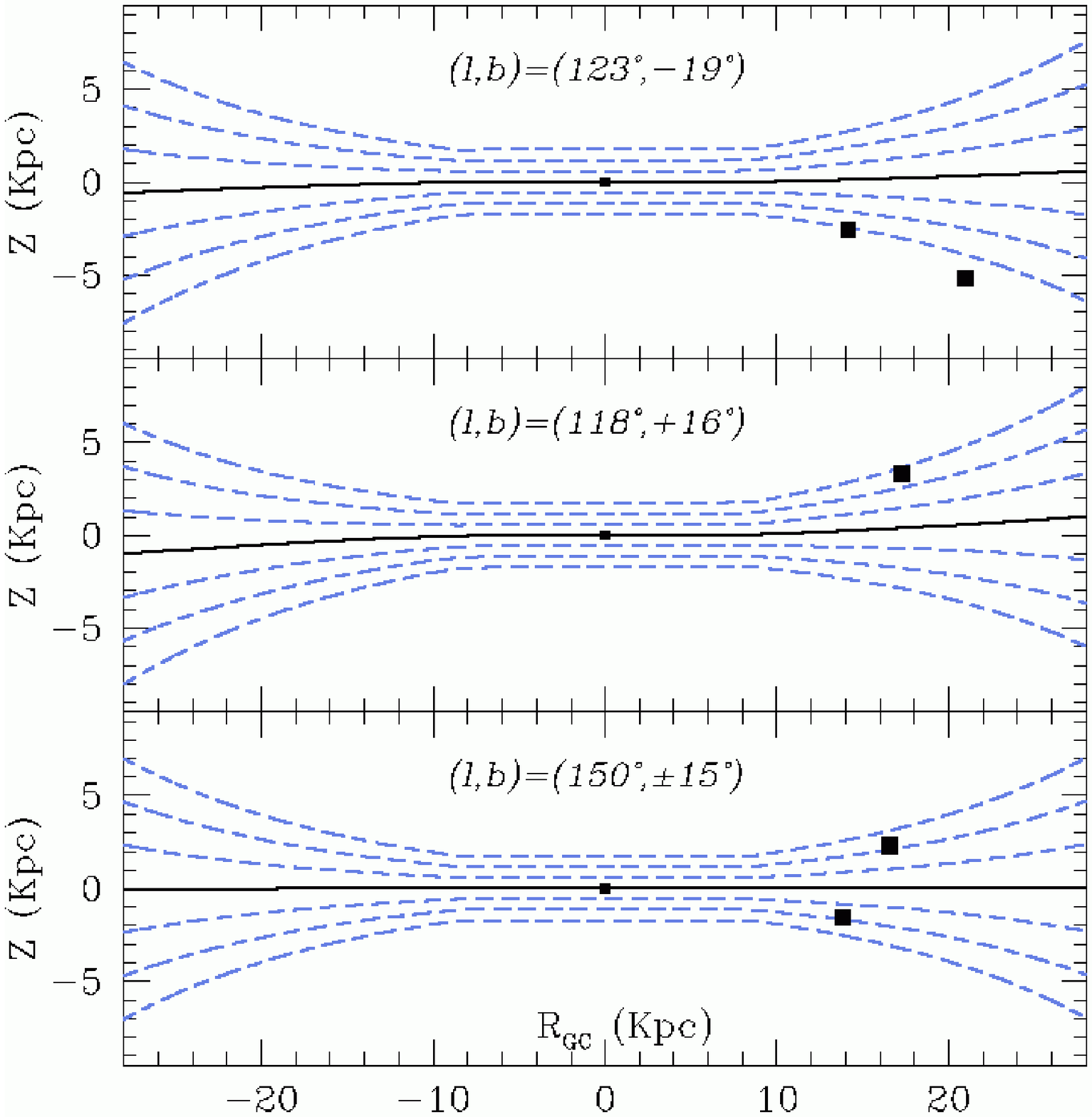}
\caption{The Same as in Fig~\ref{f_F3} but for fields where positive detections of 
the Mon.  Ring  have been reported.   For the Sloan  detections (upper
three panels) we plot detections at heliocentric distances of 7, 8, 9
and 10   kpc.   Similarly,  the  lower  three  panels   show  INT
detections  (Conn  et     al.       \cite{conn05} and   Ibata     et
al. \cite{ibata03}) plotted at the distances given  in Table~3 of Conn
et al. }
\label{f_F4}
\end{figure}
%-------------------------------------------------------------

In Fig.~\ref{f_F4} we  show  the   location  of positive Mon.     Ring
detections with respect to the warped and flared  Milky Way disk.  For
the Sloan   fields   (Newberg  et   al.  \cite{newberg02}),  we   plot
detections at heliocentric  distances of 7, 8, 9  and 10  kpc, whereas
for  the    INT  fields (Conn   et al.    \cite{conn05}   and Ibata et
al. \cite{ibata03}) we plot the detections at the specific distances
given in Table~3 of Conn et al.

Figure~\ref{f_F4} shows that all reported Mon.  Ring detections are at
locations where  the structure of the  outer warp and flare contribute
to the analysis, and  must be carefully  considered, and  at locations
where the dynamical interaction between  Monocerous and the outer disk
will be very important.
The  Mon. Ring has attributed properties  (a  radius of $15-20$ kpc, a
radial thickness of 2  kpc and a   vertical scale-height of 0.75  kpc,
Ibata  et al.  \cite{ibata03}) which   place it well  within the outer
disk defined earlier in this paper.

Of      all       the  Mon.         Ring  fields,    the        one at
$(l,b)=(150^{\circ},\pm15^{\circ})$ is perhaps the most important.
In  their INT survey, Conn  et  al.   (\cite{conn05}) argued that  the
detection of Mon. Ring structures on both sides of the Galactic plane,
at  similar Galactic  longitudes,  is the strongest  evidence that the
Mon. Ring is not an artefact of Galactic warp effects on star counts.
We    agree with   Conn et   al.     (\cite{conn05}) that   inadequate
substraction of the warp  is not a   valid explanation for Mon.   Ring
detections above  and below the  mid-plane.  We consider  here if disk
flaring provides a valid explanation  for the Mon. Ring detections  in
both hemispheres at $(l,b)=(150^{\circ},\pm15^{\circ})$.

A hint that Mon.   Ring detections are  related to disk  flaring comes
from the  field at $(l,b)=(123^{\circ},-19^{\circ})$, first studied by
Ibata et al.  (\cite{ibata03}),  who  reported a positive  Mon.   Ring
detection at $R_{GC}\sim14$ kpc. This field was re-examined by Conn et
al.  (\cite{conn05})  who   reported a  {  possible} new  detection at
$R_{GC}\sim21$ kpc.  Figure~\ref{f_F4} shows that the earlier and this
new tentative detection of Mon.  Ring features follow the expected $Z$
height of the flared MW disk.

We  are clearly far from   performing a quantitative  analysis of  the
impact of a flared disk on positive detections of Mon.  Ring.
On the one hand,  we should do our  best to decrease  the rather
large uncertainties in the distances of Mon.  Ring detections.
On the other hand, it would be interesting to have new Galactic models
in which, for  example, the radial truncation of  the  stellar disk at
$R_{GC}\sim14$ kpc is dropped, the orientation of the warp phase-angle
is updated, and new flare modeling is included.
Note  for example that the   extensive comparison between observed and
\besancon\ simulated  diagrams in  Mon.      Ring  fields (Conn    et
al.   \cite{conn05}) implicitly assume a   stellar disk truncation  at
$R_{GC}\sim14$ kpc.  
Since most of  the Mon.  Ring detections  are at Galactocentric  $>14$
kpc  it  follows that   the  comparison  with the  Galactic  model  is
inconsistent.

\section{Galactic structure to $R_{GC} =20$ kpc in the outer Galaxy: what
do we expect?}
\label{s_outer}

In the previous  Sections we have shown how  complex the outer (beyond
the Solar circle) stellar disk of our Galaxy is,  and how many intrinsic
structures can be identified in it.  We have also critically discussed
the   difficulties in determining the   stellar  disk main  parameters
(warp, flare,     scale  height)  by    studying  its   main   stellar
components. Despite these unavoidable difficulties, we have shown that
the parameters inferred from   the stellar distribution  compare  well
with  what has been  obtained from  a number  of investigations on the
distribution of   the gas, dust,  and  neutral  Hydrogen. In  order to
complete our discussion, in this section we add additional information
on what we  know about the outer  Galactic disk, and some information  of
the properties of the stellar disk in other galaxies.

%---I don't remember why this sentence is commented!!
%
%It  is an extreme  violation    of the Copernical  principle, and   so
%unlikely, to   assume  that  the   properties of the    Galaxy  change
%significantly at   the    solar Galactocentric   radius.   
%
The outer structure of galactic disks, and the Galactic disk, has been
extensively    studied for very many     years,  prior to the  current
re-discovery of its   intrinsic  interest. A  considerable  amount  of
direct  information  on the  outer  Galaxy  is available, as  is further
information by analogy from studies of other galaxies.

A superb review by   Burton (\cite{burton98}) summarizes and  explains
the subject and data,  while  presenting the observed basic  features
and results which remain  valid in  the  light of new data.   The most
extensively  studied  major  component   is  the  {\ion{H}{i}}. 
This \hi\ extends approximately smoothly and continuously to $R\sim25$
kpc   (Burton \cite{burton98} Figs   7.19,  7.20). The distribution is
significantly  warped and  flared   (Figs 7.18  to  7.23),  there  are
significant  non-circular   motions, and   the   gas is    distributed
asymmetrically in azimuth  (Fig  7.21), implying a  very significantly
lop-sided gas distribution.
The results from Burton (\cite{burton98}) were updated and extended by
Hartmann \&  Burton (\cite{hartmann97}), and  most  recently in a very
extensive  new    study  of the whole  sky    \hi\  distribution,  using
significantly  improved  calibrations and stray-light  corrections, by
Kalberla et al. (\cite{kalberla05}).  
Inspection of the Hartmann/Burton maps for velocities corresponding to
the far  outer Galaxy  (V$_{LSR}\ge+50$ km/s) shows  extremely clearly
that the outer gas layer  exists, is smooth   in velocity space  (i.e.
distance),  is  centered significantly  below   the Galactic plane,  is
lop-sided and  warped,  and is  thick. By contrast,  until the present
resurgence of interest, there have been rather few large angular-scale
studies of the stellar distribution.

% $\ell$ nice than $l$
%
%
A   very important  point to  note  in anti-center  sight-lines is the
considerable `foreground'  complexity associated  with the spiral  arm
structure, which extends well beyond  the solar circle: in particular,
there is the Perseus spiral arm just outside  the solar orbit, clearly
visible in  CO  $l-v$ maps  extending across  the entire anti-center
region  from   $l~\approx~60^{\circ}$    through  $180^{\circ}$ to
$l~\approx~260^{\circ}$.
In the  anti-center foreground, up to 2  kpc distant, especially in the
directions   towards $l=140^{\circ}-220^{\circ}$,  there  are large
angular-scale    star-forming regions, with    their associated  dust,
including Monoceros, Orion, the Taurus-Perseus-Auriga complex, and UMa
and  Cam OB-star regions.    Some of these  large-angle structures are
physically large in space (e.g. Gould's belt) and are distributed highly
asymmetrically with respect to the main Galactic Plane.

Thus  the simplest and most  robust conclusion from a direct examination
of   the   sky is that   one  does  not expect   any  simple symmetric
distribution of sources  on the sky to be  an  adequate description of
the outer Galaxy. One must select tracers  by distance, and appreciate
complex  foregrounds, and  expect  the outward  continuation of normal
spiral structure.

\subsection{\hi\ and CO surveys: a star-forming disk to $R_{GC}=20$ kpc}

Diplas  \& Savage (\cite{diplas91}), in   an early extensive study  of
outer-Galaxy  gas which  remains  valid  and  topical, conclude:   for
$12\le~R_{GC}~\le18$ kpc   the    $z$-distribution  of    the  \hi\   is
complicated, and reveals the existence of  a confined component and an
extended  component. The   confined  component exists   only within  a
Galactocentric   radius   $R_{GC}\sim18-20$   kpc, while  the extended
component can be followed in some directions to $R_{GC}\sim25-30$ kpc.
In           the         directions       of         maximum      warp
($90^{\circ}\le~l~\le~110^{\circ}$) the average  distance of the gas
away from the  the  Galactic plane is found   to reach $\sim4$  kpc at
$R_{GC}\sim24$ kpc,  and  the  flaring of   the  layer as measured  by
$z_{rms}$ increases to  $\sim3$ kpc.  In  these directions the warping
outer layer appears to be reliably traced over a  region at least $10$
kpc thick in $z$.   When viewing  from the  position of the  Sun, this
implies that  the warped and flaring  outer Galaxy extends to Galactic
latitudes as large as $25^{\circ}$.

\noindent 
This raises the interesting  question:  is the ``confined  component''
what has been re-discovered recently, the Mon. Ring ?

Nakanishi \& Sofue (\cite{nakanishi04}) completed a global analysis of
Galactic  \hi: they show extensions  of \hi\ in  the outer Galaxy well
beyond  $20$ kpc,  with evidence    of very  large scale    rotational
asymmetry  (lop-sided), so much so  that the Galaxy extends farther in
the   first  and fourth quadrants.    An   important point relevant to
current discussions is that the general outer distribution of the \hi\
is connected but is not simply smooth. Since the distances are derived
from  a kinematic   model (flat  rotation curve),  there  is  a direct
degeneracy between kinematic asymmetries - i.e. elliptical orbits, and
spatial inhomogeneities.  Outer  spiral  structure is evident, as   is
vertical warping.

We emphasize   that from all   these and the  many other   \hi\ and CO
studies    { there  is  no evidence  for  a significant
perturbation from   a   major satellite   accretion   event  currently
underway}: quite the opposite in fact,   the global approximate smoothness  of
the  \hi\ distribution  is very  difficult to  reconcile  with such  a
model.

There  are many molecular studies of  the outer Galaxy indicating that
the gas undergoes ``normal''  outer galaxy rates and distributions  of
star formation, again supportive of a largely steady-state system.  
An example is the Brand \&  Wouterloot (\cite{brand95}) study of outer
IRAS    sources,  which  are star-forming    molecular   clouds out to
$R_{GC}\simeq20$   kpc.   They collect    data  from  several  earlier
studies. Snell et al.  (\cite{snell02}) provide a valuable overview of
both the Galactic data and studies  of other galaxies, { concluding
that outer  Galaxy molecular cloud properties  and star formation is a
natural extension of the inner Galaxy properties}.

We   note a  study  by   Freudenreich  (\cite{freud96}) of  the  DIRBE
observations of outer  galaxy dust, which  reports a detection only to
$R_{GC}\simeq12$ kpc.  It  is unclear why this  study  fails to detect
the farther  Galactic   molecular  material, but  this    paper is the
anomaly, all the other studies agree rather well.

A major CO  study  by Dame, Hartman   \& Thaddeus (\cite{dame01})  and
recent extension   to  higher     latitudes  by Dame     \&   Thaddeus
(\cite{dame04}) shows extended    CO emission   at latitudes up     to
$50^{\circ}$ in the anti-center region,  with this gas probably  being
associated  with star formation in  Taurus-Auriga.  Their Fig.~3 shows
very well the outer  molecular gas at   velocities up to $+100$  km/s,
corresponding to very large Galactocentric distances.

The most  recent high-sensitivity southern  \hi\  survey, Dickey  et al.
(\cite{dickey04})  shows that the  \hi\  warp continues to more-negative
$z$-distance as distance increases to a minimum between $R_{GC}=13-16$
kpc, with  the location a  function of longitude through this distance
range, and then returns to near $Z=0$ by  $R_{GC}=20$ kpc. Beyond $20$
kpc and  out  to $30$ kpc the    \hi\ distribution `is dominated'   by a
density enhancement  which extends  coherently over very  wide angles,
and  may be  a coherent  extension    of the  Perseus spiral arm   for
longitudes $l~\approx~220^{\circ}-325^{\circ}$.
This very large structure, if indeed it is coherent, may be similar to
the  spiral-like  structures in  \hi\ extending well  beyond the visible
optical  disk seen  in   several other galaxies, with  associated  star
formation    and apparently   normal     structures, by  Ferguson   et
al.  (\cite{ferguson98}) and  the  several   other examples   recently
observed by UV analysis.

As  an example of  studies of external   spirals, we note the complex,
asymmetrical,  inflowing \hi\  seen in   NGC2403.  Fraternelli et  al.
(\cite{fraternali02}) show this  is  a normal, non-interacting  spiral
galaxy, suggesting  that such \hi\ features may  be common  among spiral
galaxies and that, perhaps, they have not been detected yet because of
the low sensitivity of previous surveys.

\subsection{Expected kinematics}

Bailin \& Steinmetz (\cite{bailin03}) note that warps are very common,
so much so that  perhaps  50\% of  galaxies  have them  (Reshetnikov \&
Combes (\cite{reshetnikov98}).   But,  warps are not  long-lived (e.g.
Binney \cite{binney92}), so they must be generated in some way.

Hence, if a warp is kinematically driven, there is no reason to expect
simple geometry or  kinematics.    The actual kinematic  pattern  will
depend on  what is  the cause, which   is not clear.   Possible causes
include the cumulative effects  of many satellites (not necessarily on
decaying orbits), though in the case of the Galaxy it  may just be the
LMC which induces a large-scale reaction from the DM halo (Weinberg
\cite{weinberg98},    Tsuchiya         \cite{tsuchiya02},     but  see
Garc{\'{\i}}a-Ruiz, Kuijken \&  Dubinski \cite{garcia02}) or maybe  by
Sagittarius  (Bailin \& Steinmetz \cite{bailin03}).
Bailin  \&  Steinmetz emphasize that  {a  warp generated by interaction
with a satellite is  a dynamical warp, so that  one does not expect to
see  stationary  or   uniformly precessing kinematics}.     A transient
induced warp moves outward at the group velocity of bending waves in a
stellar  disk,  which  scales as $\pi~G\times~\Sigma/\Omega$, for
surface density  $\Sigma$ and rotational  velocity $\Omega$ (Hofner \&
Sparke \cite{hofner94}) or $\sim30 $kpc/Gyr near the Sun.

Newberg \& Yanny (\cite{newberg05}) suggest significant triaxiality in
the large scale structure of the outer  galaxy star counts. This is for
distances of  $3-16$ kpc exterior  to  the Sun.   This follows earlier
suggestions  by Roberta  Humphreys.   
Note that they see similar triaxiality  and coherent radial motions in
the  local old  disk, a  pattern perhaps consistent  with the combined
perturbations on circular  orbits  due  to  spiral structure (cf    de
Simone,   Wu \&   Tremaine  \cite{simone04}) and    effects  due to  a
bar-shaped perturbation in the disk/bulge  potential (cf Babusiaux  \&
Gilmore \cite{babu05}).   In  any  event, simple  circular  orbits are
neither natural nor expected.

\subsection{Expectations from other galaxies}

Star formation  in very  outer disks,  beyond  two $R_{25}$, has  been
detected  from  H-alpha  imaging in  several  galaxies  (e.g.  NGC628,
NGC1058, NGC6946 Ferguson et  al.  \cite{ferguson98}).  Thilker et al.
(\cite{thilker05}) GALEX imaging the outer  disk of M83, shows current
star formation at up to three $R_{Holmberg}$  in the warped outer disk
of M83. The  extreme patchiness of the very  outer disk is evident  in
Fig.~1 of Thilker et al., { yet there is no suggestion that what is
observed is anything other than  star formation in  a disk, albeit  at
very large Galactocentric distances}.
The GALEX images show star formation, and the  existence of both young
and   intermediate-age open clusters, well   beyond  an apparent  disk
`edge'.    The authors note the   location  of the young  star-forming
regions ``preferentially on local maxima or filaments in the structure
of the warped  \hi\  disk'',  with these  complexes superimposed   on an
extended field stellar population.

Gentile et al. (\cite{gentile03}) looked  at ESO123-G23, an apparently
thin edge-on galaxy with a warp in the optical,  and saw \hi\ warped gas
up to $15$ kpc  from the optical major  axis, implying a  warp through
$30^{\circ}$.

An interesting specific study, which  reminds us that the dynamics  of
outer galaxies may not always require current  active mergers, is that
by Fraternali  et al. (\cite{fraternali02})  of NGC2403.  They studied
the spiral galaxy NGC2403  with the VLA,  and discussed the properties
of  the extended, differentially rotating \hi\  layer with its \hi\ holes,
spiral structure, and outer warp.
In addition, their new data reveal the  presence of a faint, extended,
and kinematically anomalous component.  This shows up in the \hi\ line
profiles as extended  wings of emission  toward the systemic velocity.
In the central regions these wings are very broad (up to 150~km/s) and
indicate large  deviations  from circular motion.  They  separated the
anomalous  gas component from  the cold disk  and obtained  a
separate velocity  field and a separate rotation  curve for each.   The mass of
the anomalous component is 1/10 of the total  \hi\ mass.  The rotation
velocity of the anomalous  gas is $25-50$  km/s lower than that of the
disk.  Its velocity field has non-orthogonal major and minor axes that
they interpret as  due to  an  overall inflow motion  of $10-20$  km/s
toward  the center  of the galaxy.   The picture  emerging from  these
observations is  that of a cold  \hi\  disk surrounded by a  thick and
clumpy \hi\ layer characterized  by slower rotation and inflow  motion
toward the   center.  The origin   of  this  anomalous  gas  layer  is
unclear. It is likely, however, that it is related to the high rate of
star formation in the  disk of NGC2403  and that its kinematics is the
result of a galactic fountain-type of mechanism.

In a related study, Fraternali  et al. (\cite{fraternali04}) show that
thin `cold'  \hi\ disks are often surrounded  by thick layers (halos) of
neutral gas with anomalous kinematics.  They present results for three
galaxies  viewed at different   inclination angles: NGC891  (edge-on),
NGC2403  ($i\simeq60$), and NGC6946   (almost face-on).  These studies
show the presence of halo gas up to distances  of $10-15$ kpc from the
plane.  Such gas  has a mean rotation  $25-50$ km/s lower than that of
the gas in  the plane, and some complexes  are  detected at very  high
velocities, up to $200-300$ km/s.  The nature  and origin of this halo
gas are poorly understood.  It can either be  the result of a galactic
fountain   or of  accretion  from the    intergalactic medium.   It is
probably  analogous to some of the  High Velocity Clouds (HVCs) of the
Milky Way.

The overall picture from these many studies is that the outer parts of
galaxies   are complex,  asymmetric   and time-dependent.   While  not
understood, ongoing mergers of  significant companions seem not  to be
the primary explanation or cause at work.

\section{Discussion and Conclusions}

We have used  the 2MASS all-sky catalog to  probe two important  outer
Milky Way parameters; the Galactic warp and flare.
We show that the Galactic warp can fully explain the detection of the
Canis Major over-density.  We also show that the Galactic flare of the
outer disk will be apparent at those places where there are detection
of Monocerous Ring over-densities.
Thus, the results presented in  this paper provide a reminder
that disentangling the accretion history of the outer disk from its
complex and time-dependent dynamics remains a challenge.
Our results add to the well-known evidence  of the complexity of outer
Milky Way structure.

{\em \bf The Galactic warp}: using 2MASS red clump and red giant stars
we  first  demonstrated the  dependency  of the   warp parametrization
(amplitude and phase-angle) on the contamination status of the adopted
stellar tracer.  Red giants are proved excellent  tools in probing the
outer most parts of the Milky Way.
The   detection of a  global  warp  signature   by RGB  stars  at mean
heliocentric  distances of $R_{\odot}=2.8$, $7.3$  and $16.6$ kpc, has
helped us understand the following:

\begin{itemize}
\item  The warp phase-angle is oriented in such a way that at fixed
distances from the Sun we observe an asymmetric warp (relatively stronger
maximum amplitude  in  one  hemisphere   with  respect  to  the  other
hemisphere).   For  the $R_{\odot}=2.8$,   $7.3$ kpc RGB  samples, the
Southern   hemisphere warp maximum shows    twice  the amplitude  of
that  in  the Northern hemisphere.   A symmetry of the  warp
maximum  is obtained only  for the $R_{\odot}=16.6$  kpc, at which the
dependency on the phase-angle is weakened;
\item The warp phase-angle ($\phi\simeq+15^{\circ}$) is close to the
orientation of the Galactic bar;
\item The detection of the stellar warp in the RGB sample at a mean
distance $R_{\odot}=2.8$ kpc proves  that the  warp starts within  the
solar circle;
\item  The detection of the stellar warp in the RGB sample at a mean
distance $R_{\odot}=16.6$ kpc proves the presence of an extended Milky
Way stellar population  out  to $R_{GC}\sim24$ kpc.  This  is in clear
contradiction   with  the assumption  of a   radial  truncation of the
stellar disk at $R_{GC}\sim14$ kpc.   
The  detection of an extended stellar  disk is  confirmed by parallel,
and independent studies reporting: (i) recent star formation, traced
by molecular clouds, out  to $R_{GC}\sim20$ kpc;  and (ii) an extended
\hi\ spiral arm in the Third and Fourth quadrants;
\item For the 3 RGB samples, the Southern  hemisphere  warp    maximum
is found around $l\simeq240^{\circ}$, we  therefore establish the {\em
identity of the CMa center with the warp maximum}. 
\item The derived stellar warp shows excellent agreement (in both
the phase-angle and amplitude of the maximum) with that of the gaseous
and   interstellar   dust   distribution.   This    resolves 
conflicting evidence of  amplitude differences between the  3 Galactic
components. The origin of the different warp signatures is due
to foreground contamination and distance resolution of the adopted tracer;
\end{itemize}

These results  do not answer the  fundamental question of {\it what is
the driving mechanism  of the warp?}  There is no general agreement on
which     process,    or     processes,      generate      warps  (see
Castro-Rodr{\'{\i}}guez et al. \cite{castro02}).
Satellite  accretion remains one  among many scenarios  put forward to
explain warps.   Other possibilities include:  cosmic infall,  bending
instabilities,     intergalactic   magnetic   field,  accretion     of
intergalactic medium (see also Olano  (\cite{olano04}) for a  possible
warp-HVCs-Magellanic Clouds connection).
Any proposed scenario in a special case should also account for the
fact that warps are a ubiquitous property of most, perhaps all, disk
galaxies (S{\'a}nchez-Saavedra et al. \cite{sanchez03}).

{\em \bf The Canis Major Overdensity}: We have  shown here that CMa is
the  southern hemisphere maximum stellar  warp.  We also note that all
CMa properties (measured so  far) point to the over-density reflecting
a normal  warped disk population  rather  than that of  a cannibalized
dwarf galaxy.  Among these we note:
\begin{itemize}
\item the  absence of   any  peculiar  radial velocity signature   (see
Appendix.A);
\item a possible shortage of an old ($\ge$9 Gyr) and metal poor stellar
population (see Appendix.B);
\item the absence of any peculiar proper motion signature (see
Appendix.C) and a rather problematic orbital motion\footnote{The
near circular orbit of CMa is hard to account for in accretion models,
usually preferring elliptical orbits.  However, Crane et al.
(\cite{crane03}) pointed out that this particular issue cannot rule
out the accretion scenario since the tidal disruption of satellites
depends on whether these satellites {\em were originally born} with
near-circular orbits (Taffoni et al. \cite{taffoni03}).}; and
\item a rather high line-of-sight extent (see Appendix.D).
\end{itemize}

{\em \bf  The Galactic flare}: having characterized  the warp, we also
derived the Galactic flare for the 3 RGB samples.
Unable to kinematically discern  thick from thin disk populations, the
retrieved variation with $R_{GC}$ of the disk thickness shows a rather
constant scale-height ($\sim0.65$     kpc) within $R_{GC}\sim15$ kpc. 
Further out, the mean scale-height increases gradually reaching 
$\sim1.5$ kpc at  $R_{GC}\sim23$ kpc. 
Whereas this trend refers to a mixture of thin-thick disk populations,
for the RGB  sample at $R_{\odot}\sim2.8$ kpc  (dominated by thin disk
populations) we derive a  mean scale-height of  $\sim0.35$ kpc that  is
consistent with typical values for the thin disk.

Overall, we  trace the  stellar  disk  flaring out  to  $R_{GC}\sim23$
kpc. Thus, we add  further and final  evidence on the inconsistency of
the disk radial truncation hypothesis, at $R_{GC}\sim14$ kpc.
Indeed,  the  detection of stellar disk   warping and  flaring  out to
$R_{GC}\sim23$ kpc is the  key answer to  the observed drop in stellar
density near the Galactic plane.

\begin{acknowledgements}
We  are indebted  to the  anymous  referee the useful suggestions that
improved   the overall quality  of  the paper.  YM,  GP,   LRB and FdA
acknowledge the support by the Italian  MIUR under the program PRIN03.
YM thanks  Marcel Clemens and Melissa Evans  for a  careful reading of
the manuscript, and H.I.N.A.S for all the support.

This publication makes use  of data products from  the Two Micron  All
Sky  Survey,   which is   a joint   project   of  the   University  of
Massachusetts   and  the     Infrared    Processing   and     Analysis
Center/California  Institute  of  Technology, funded   by the National
Aeronautics  and  Space   Administration   and the   National  Science
Foundation.    We also  make   use of  the    WEBDA  database and NASA
ExtraGalactic Database.

\end{acknowledgements}

%----------------------------------###
\appendix
\label{app1}

%\section{The radial velocities of CMa}
\section{The  absence of  peculiar  CMa radial velocity signature}

In ``Why the Canis Major over-density is not due to the Warp'', Martin
et al.  (\cite{martin04b}) presented radial  velocities of M-giants at
the center of CMa.  The overall  distribution of radial velocities was
strongly  characterized  by  a   double-peak  morphology.  Martin   et
al. used this distribution to argue that the peak at $V_{rad}=109\pm4$
km/s is   to  be attributed to  an   extra  population, CMa,  that  is
distinguishable from typical   disk stars peaking  at $V_{rad}=61\pm4$
km/s.

A recent analysis  by the same  group (Martin et al.  \cite{martin05})
of  more than 1500  RGB and RC stars,   showed that the bimodal radial
velocity  distribution   was  {\em  artificially  produced by template
issues}.  The new $V_{rad}$ distribution is now broad; there is no
peculiar signature of CMa stars and these are indistinguishable from
surrounding disk stars.
Nevertheless, Martin et al.  proceeded and have identified
a ``{\em peculiar distance-radial   velocity relation that  is  unlike
that expected from  thin or thick  disk stars}''.  That  is, while CMa
stars show a mean radial velocity of  $72\pm7$ km/s at $R_{\odot}=$5.5
kpc,  the  mean   $V_{rad}$   increases to about  $114\pm2$    km/s at
$R_{\odot}=$8.5 kpc.

To date, this ``peculiar''  $R_{\odot}-V_{rad}$ relation is considered
the  strongest kinematic evidence  standing  in favor of an  accretion
scenario.   However, as  we  shall  demonstrate  below,  this relation
reflects nothing more than the Galactic differential rotation.
%

%------------------------------------------one column figure
\begin{figure*}
\centering\includegraphics[width=13cm,height=13cm]{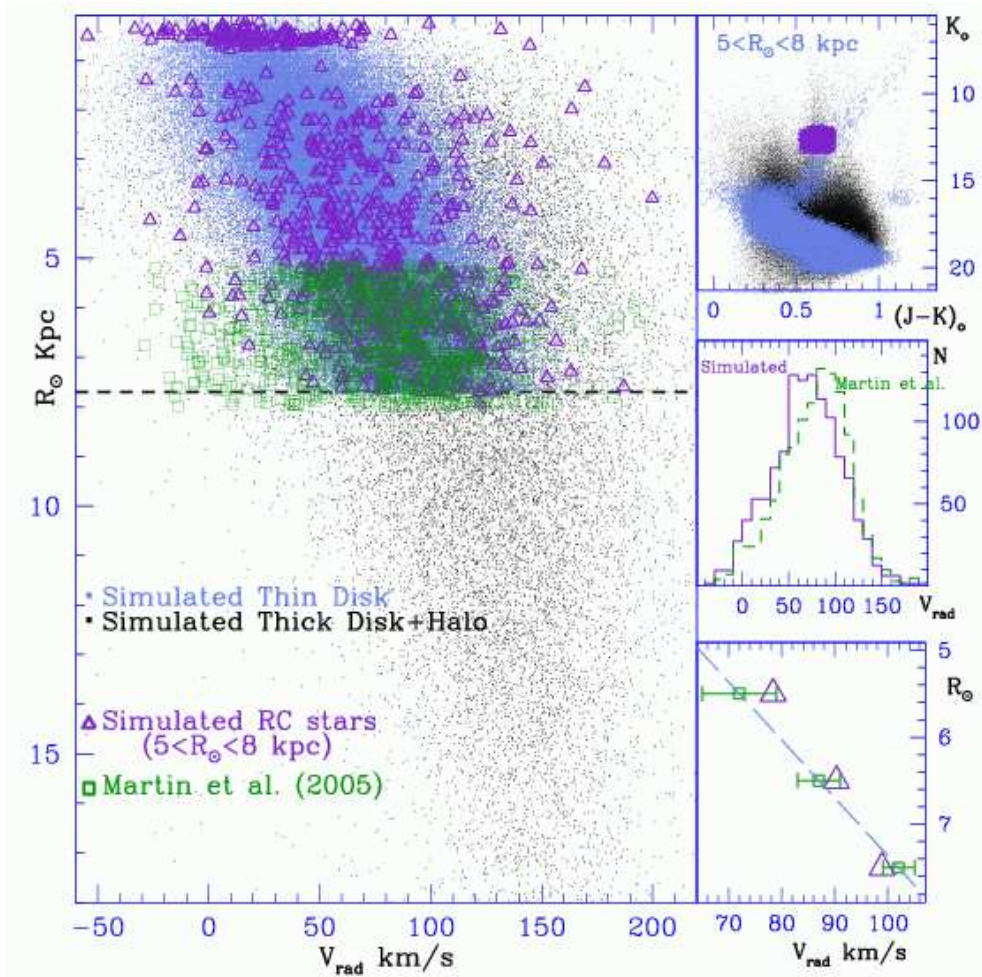}
\caption{Explaining the ``peculiar'' $R_{\odot}-V_{rad}$
relation found for CMa stars as due to Galactic differential rotation.
The left   panel shows   a  \besancon\  simulated  $R_{\odot}-V_{rad}$
distribution for stars at the CMa center.
Grey  dots  highlight  the   modeled  {\it  Thin  Disk} truncation  at
$R_{\odot}=7.7$ kpc (marked  by a horizontal  line).   Black dots plot
the {\it Thick Disk} and {\it Halo} stars.
Open triangles show  the  $R_{\odot}-V_{rad}$ distribuition  for  {\em
simulated} stars extracted  from the synthetic color-magnitude diagram
(upper right panel), and falling within a box delimiting the RC region
between $5\le R_{\odot} \le8$ kpc.
The open squares  are the {\em  observed-RC} sample from Martin et al.
(their Figure 6).
Right middle panel shows the distribution of  radial velocities of the
Martin et al.  {\em  observed-RC}  sample (dashed histogram), and  the
scaled  distribution   of    {\em    simulated-RC} stars   (continuous
histogram).  
The   right  lower  panel   shows  the  mean   $V_{rad}$ of   the {\em
dwarfs-decontaminated  and  RC-simulated}    sample (open   triangles)
estimated in three distance intervals.
The open squares plot the {\em observed} mean $V_{rad}$ values for CMa
RC  stars (see  Table   2 of  Martin  et  al.) in  the   same distance
intervals. The error bars are from the same table.
The Dashed line is a linear fit to the Martin et al. 
mean values;  which is fully compatible with  recent estimates  of the
$A$ Oort constant being $A=17.23\pm0.35$. }
\label{f_vrad}
\end{figure*}
%-------------------------------------------------------------

To shed more  light on this ``peculiar'' $R_{\odot}-V_{rad}$ relation,
we combine the results obtained by Martin et al.  (their Table 2) with
a kinematic simulation from the \besancon\ Galactic model.
In  the     left   panel  of  Figure   (\ref{f_vrad})   we    show the
$R_{\odot}-V_{rad}$ distribution for {\em  all simulated stars} in the
simulated field centered at ($l,b$)=($240^{\circ}$,$-8^{\circ}$).
We remind the reader that the  \besancon\ model imposes a stellar disk
truncation  already   at $R_{\odot}=7.7$   kpc     ($R_{GC}\simeq14.0$
kpc)\footnote{As  demonstrated in  Section~\ref{s_flare}, the  results
presented  in  this paper  confirm  earlier  suggestions that  a  disk
truncation hypothesis  at  $R_{GC}\simeq14.0$  is  neither needed  nor
justified.}, more or less at the CMa distance.
Thus, at about  $R_{\odot}=7.7$  kpc and beyond, one  {\em cannot
quantitatively}  compare the observed  kinematics  with the incomplete
\besancon\ simulation.
To highlight   the abrupt truncation of  the  simulated Galactic disk,
{\it Thin Disk} stars  within $R_{\odot}=7.7$ kpc  are plotted as grey
dots, while those further  out ({\it Thick Disk}  and {\it  Halo}) are
plotted as black dots.
Before going  into  more detail,   and prior  to  any selection    or
analysis,  one   already    sees  that    there   is  an    intuitable
$R_{\odot}-V_{rad}$ trend in the simulated data, similar to that found
in the Martin et al.  analysis for CMa stars.

The RC  sample of Martin et al.    (their Fig.~6) is presented  in the
left panel (open squares).
In order to compare this  {\it observed-RC} sample with an appropriate
counterpart from  the  \besancon\  simulation   (a {\it  simulated-RC}
sample), we extract  RC stars from  the one  square degree, dereddened
$K_{\circ}$, ($J-K$)$_{\circ}$ diagram.
The simulated color-magnitude diagram is displayed in the upper
right panel.
In   grey  symbols we highlight    all  simulated stellar populations
between $5\le R_{\odot}\le8$ kpc.
The selection box, where  we extract {\it simulated-RC}  stars between
$5\le R_{\odot}\le8$ kpc is shown as grey rectangle area.
We emphasize  that this {\it simulated-RC}  sample, as that studied by
Martin et al., is subject to foreground contamination.

The advantage  of using simulated data  is that one can easily isolate
the foreground dwarf contamination from the {\it simulated-RC} sample,
by means of the simulated stars distance entries.
All stars included in the {\it simulated-RC} sample are plotted in the
left   panel (open triangles),    showing clearly the contamination by
local dwarfs within $\sim1$ kpc.
Between $5.5$ and $8.0$ kpc {\it simulated-RC}  stars overlap with the
Martin et al.   observed sample (a   colored version of the  figure is
more appropriate for the disentangling).

The {\it middle right  panel} shows the radial velocities distribution
of the Martin  et al.   {\it  observed-RC} sample (dashed  histogram).
The  distribution of  {\it  simulated-RC} stars as  extracted from the
\besancon\ simulation is plotted as a continuous histogram.
We make use of  the Martin et al.   (\cite{martin05}) estimate of  the
$V_{rad}$ intrinsic dispersion  of disk stars  at the CMa distance and
add this value (via a Gaussian distribution) to the {\it simulated-RC}
stars.
The model   histogram has been scaled  to  fit the observed histogram:
other  appropriate  scaling is  not  possible  since  the   Martin  et
al. (\cite{martin05}) selection function of RC stars is unknown.
The similarity of the two distributions is remarkable, we note however
that the histogram of {\it simulated-RC} stars shows a slightly higher
distribution (with respect to  the  {\it observed} sample)  for  stars
with  $V_{rad}<80$  km/s,  and a  lower distribution  for $V_{rad}>80$
km/s.
The later feature  can be explained as  being due to the truncation of
disk  population in  the \besancon\    model at $R_{\odot}=7.7$   kpc:
simulating  more distant populations  will result in  more stars only
having $V_{rad}$ higher  than 80 km/s.  On  the other hand, the higher
distribution of {\it simulated-RC} stars  at $V_{rad}<80$ km/s can  be
due to a  stronger foreground dwarf  contamination in  the input model
(note in fact  the discrete and  distinguishable distribution of local
dwarfs within 1 kpc).
The two distributions    remain, however, very   similar and  this  is
further demonstrated  in the  {\it   lower right panel} where   we (i)
exclude  nearby dwarfs from  the  {\it simulated-RC} sample, (ii)  use
this {\it cleaned simulated-RC} sample to  estimate the mean $V_{rad}$
in 3  distances  intervals  [$5\le~R_{\odot}~\le6$, $6\le~  R_{\odot}~
\le7$ and $7\le~R_{\odot}~\le8$ kpc]; and
(iii)  compare the {\it cleaned and simulated}  $V_{rad}$  values with those
derived  by Martin et     al  (\cite{martin05}), in the 3     distance
intervals.

We emphasize the fact that the mean CMa $V_{rad}$ values  derived in the 3
distance intervals (Martin et al \cite{martin05}, open squares in the
lower right panel)  are based  on what  they  call ``CMa RC  sample'',
which they disentangle from the ``contaminating population''.
Keeping in mind all the uncertainties associated with the
\besancon\ model, one clearly sees that the {\it cleaned simulated-RC}
sample (open  triangles in the lower  right  panel) shows, {\em within
1$\sigma$}, the   same  $R_{\odot}-V_{rad}$ trend found   by Martin et
al.  for a   cleaned  CMa RC sample. 

%--------
One last piece of evidence which demonstrates that the ``{\em peculiar
CMa observed $R_{\odot}-V_{rad}$  trend}'' is nothing but the  imprint
of Galactic differential rotation of normal disk stars is shown in the
lower panel of Fig.~\ref{f_vrad}.
Indeed, a linear  fit of the three points   reported by Martin et  al.
(\cite{martin05})   yields $R_{\odot}=0.067\times~V_{rad}$. Using  the
Oort equation:

\begin{equation}
V_{rad}=A~\times~R_{\odot}\times~sin(2\times~l^{\circ})\\
\label{e_oort1}
\end{equation}

\begin{equation}
V_{rad}=A~\times~sin(2\times~240^{\circ})\times~R_{\odot}=(1/0.067)\times~R_{\odot}\\
\label{e_oort2}
\end{equation}

one easily  gets    $A=17.23\pm0.35$, fully  compatible  with   recent
estimates of the $A$ Oort constant (see Dehnen \cite{dehnen98}).
Obviously, one has to consider the CMa large heliocentric distance and
location  with respect  to the   Galactic  disk, however, the   linear
relation is  really surprising, and does  not leave room for any other
easy interpretation.

Given that (i) the bimodal $V_{rad}$ distribution (Martin et al.
\cite{martin04b})     was  due to an     artifact in their
reduction   procedure,   and     (ii)  that  the    claimed   peculiar
$R_{\odot}-V_{rad}$ relation  (Martin et al.  \cite{martin05}) is well
reproduced by  a     self-consistent  kinematical  model,  and    most
importantly, seems to be nothing more than the imprint of differential
Galactic rotation.
We therefore conclude that the {\em  stellar populations of CMa do not
show any peculiar or  distinguishable $V_{rad}$ signature with respect
to normal disk stars}.

%\section{On the  ancient and metal-poor population of CMa }
\section{The possible shortage of an old ($\ge$9 Gyr) and metal poor
CMa stellar population}

All studied dwarf galaxies  are  known to  possess an ancient  stellar
population that  is traced by  the  detection of  either $\ge9$ Gyr RR
Lyrae  variables  and/or red/blue horizontal   branch stars of similar
age.  Dwarf galaxies in the Local Group show no exception to this rule
[Mateo   \cite{mateo98},  Tolstoy     \cite{tolstoy00},     Grebel
\cite{grebel01}]\footnote{Izotov \& Thuan (\cite{izotov04}) proposed
a scenario in which young ($\sim500$ Myr) galaxies like \zwi~are still
being   born in  today's  universe.   Nevertheless,  in Momany et  al.
(\cite{momany05b}) we showed that \zwi~can be much  older and that, at
the moment, there is no photometric  evidence of the existence of young
galaxies.}. To this adds the most recent entry, Ursa Major (Willman et
al.   \cite{willman05}), which shows a clear  clump of blue horizontal
branch stars.
Those in favor  of an extra-Galactic  accretion scenario of CMa  often
highlight the similarities (e.g.  mean  metallicity, mean mass,  M$_V$
etc) between the  CMa stellar  populations  and the Sagittarius  dwarf
spheroidal.    As a   consequence, it  is   particularly  important to
investigate the presence of     an old ($\ge9$ Gyr)  and    metal-poor
population of CMa.

Needless to   say, an old and  metal-poor  population is  necessary to
metal  enrich successive generations   of stars, otherwise one  cannot
explain  how a metal-rich   galaxy  ([Fe/H]$\ge-1$) could have   built
metal-poor   clusters.      
The Sagittarius dwarf fulfills this requirement; i.e. besides being on
average a metal-rich galaxy, it possesses  a metal-poor component that
is traced by RR Lyraes and blue horizontal branch stars (see Monaco et
al.    2003 and references   therein), and possibly extreme horizontal
branch   stars (Momany et   al. \cite{momany04a}).   This  explains the
progeny of metal-poor clusters like Terzan 8 and M54.
Studying the Sagittarius stellar  populations outside the tidal radius
of M54, Monaco et al.  (\cite{monaco03}) proved that the group of blue
horizontal branch  stars identified in  S341+57-22.5 by Newberg et al.
(\cite{newberg02}), are fully   compatible with  the  blue  horizontal
branch  of Sagittarius.   This shows  that  a  stream, {\em  narrow by
definition}, originating from a disrupting  dwarf is likely to leave a
footprint of {\em all} its stellar populations (both red and blue).

\begin{figure}%[ht]
\centering\includegraphics[width=8cm,height=8cm]{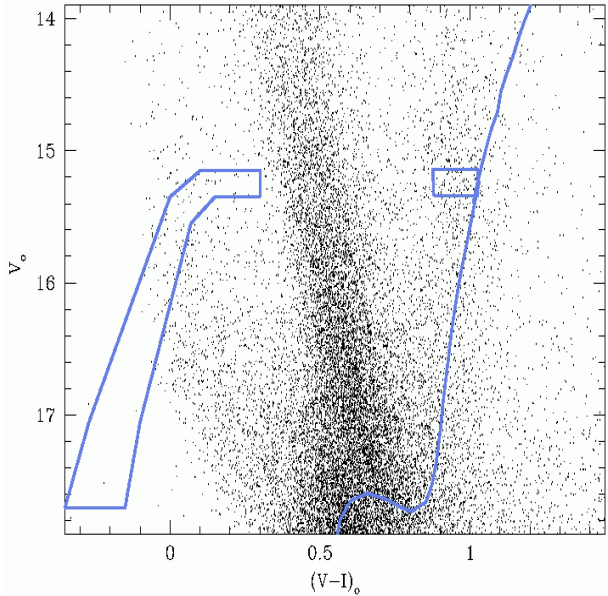}
\caption{A 1$^{\rm o}$ dereddened color-magnitude diagrams of the
core of CMa. Over-plotted is a 6.3 Gyr $Z=0.004$ ([Fe/H]$\simeq-0.76$)
isochrone   from The  Girardi  et al.     (\cite{leo02}) library.  The
assumed distance  modulus is  ($m-M$)$_{\rm  o}=14.5$.  Each star  was
dereddened    according    to its  position       in  the Schlegel  et
al. (\cite{schl98}) reddening maps. Also plotted are the red clump and
blue  horizontal branch location  of  the  Sagittarius, as derived  by
Monaco et al. (\cite{monaco03}).}
\label{f_bhb}
\end{figure}

Despite the   fact  that contamination    can   seriously hamper   the
identification of old blue  horizontal branch stars near the  Galactic
plane, we note that the CMa over-density  has not been associated with
a clear identification, or even a hint, of  old blue horizontal branch
stars.
Figure~\ref{f_bhb} displays  a wide-field color-magnitude diagram
of the CMa center, based on archival ESO/2.2m telescope data. The data
has been reduced    and calibrated following the  standard  techniques
presented in (Momany et al. \cite{momany01}).
Upon  this color-magnitude diagram we over-plot  the mean location of
red  clump and blue horizontal branch  stars in  the Sagittarius dwarf
(as derived  in Monaco et  al. \cite{monaco03}), shifted  to match the
CMa red clump population.  No hint of  an over-density of stars at the
expected blue horizontal branch location can be seen.
Instead, the distribution of stars between $14.0\le~V_{\rm o}~\le16.5$
and   $(V-I)_{\rm   o}\le0.3$  smoothly   increases   towards  fainter
magnitudes.
We  remind the reader that   at $V_{\rm o}\simeq16.0$, and in  between
$0.0\le~(V-I)_{\rm o}~\le0.5$  there lies a  population  of young $\le
100$ Myr main sequences stars which  belong to the Norma-Cygnus spiral
arm, as demonstrated by Carraro et al.  (\cite{carraro05}) by means of
$UBV$ two-color diagrams.  This population is intermediate between the
disk     MS  stars   [oblique  sequence   extending   from $(V-I)_{\rm
o}~\simeq0.2$ at $V_{\rm o}\simeq12.0$ to $(V-I)_{\rm o}~\simeq0.6$ at
$V_{\rm o}\simeq18.0$] and the blue plume population at bluer colors.
An eventual over-density due to the presence of blue horizontal branch
star ($7,000\le~T_{\rm eff}~\le11,000$K)  should take place in between
the  disk main sequence  and blue  plume,  but this  region lacks  any
significant over-density.
A similar indication is found  in the optical color-magnitude diagrams
of  Mart{\'{\i}}nez-Delgado      (\cite{delgado05})     and Bellazzini
(\cite{luna05}).

%------------------------------------------one column figure
%-------------------------------------------------------------

Recently, Mateu et al.   (\cite{mateu04}) reported on a preliminary RR
Lyrae search  in an area of  8.3 square degrees  at the center  of the
CMa. Five RR Lyraes were found at  heliocentric distances of less than
6.5 kpc.  Assuming the most recent CMa  distance range (6.2-8.2 kpc by
Bellazzini et al.  \cite{luna05}) the 5 variables with a mean distance
of $5.6$ kpc are outside the main body of the CMa over-density.
Yet, as noted by the  Mateu et al., the  detection of $5$ variables is
higher  than what  would  be expected in  the  same volume of the {\em
Galactic Halo} ($\sim$1 variable).  
Unfortunately,  a direct comparison  with  the expected number of {\em
Disk} RR Lyrae is difficult since the distribution (and density) of RR
Lyrae stars   in the {\em  thick disk}  is  not known.   Moreover, one
should add the current uncertainty in  (i) disentangling Disk and Halo
RR Lyrae at low Galactic latitudes, and  (ii) the disk density profile
as a function of $R_{GC}$.  As demonstrated  thoroughly in this paper,
the disk   radial  extension does  not  show  an abrupt truncation  at
$R_{GC}\sim14$  kpc, and  this might  increase the  number of expected
variable stars.

We therefore conclude that there is no evidence which demonstrates the
presence of an old, blue horizontal branch population CMa. At the same
time, RR  Lyrae surveys  leave this  issue rather  2open,  highlighting
intrinsic  difficulties (expected Galactic  contamination and relative
vicinity of CMa) of similar studies.
More data is needed to draw a firmer conclusion.

\section{The  negative vertical velocity of the Canis Major}

Dinescu   et al.  (\cite{dinescu05})  have   recently  re-measured the
absolute proper motion at the center of  the Canis Major over-density,
which was first determined by  Momany et al.  (\cite{momany04b}  using
UCAC2    catalogs\footnote{See    however     Momany    \&      Zaggia
(\cite{momany05a}) reporting  the   presence   of  possible systematic
errors in the UCAC catalog.}).
The authors  find that while CMa  has an in-plane rotation (similar to
the mean of thick disk  stars) it  shows significant 3$\sigma$  motion
perpendicular to the disk.  In turn, they find this inconsistent (at a
7$\sigma$  level) with  the expected   motion  of  the warp  at  these
Galactic locations    (estimated   in   Drimmel,  Smart  \&   Lattanzi
\cite{drimmel00}).
This incompatibility    lead Dinescu   et  al.   (\cite{dinescu05}) to
conclude that CMa is part of a satellite galaxy remnant.

To resolve the claimed incompatibility we note that the Dinescu et al.
(\cite{dinescu05}) conclusions  rely on two points.   The first is the
assumption  that  their selected sample  [used   to measure the proper
motion and consisting of ``{\it likely  main sequence stars}'' or blue
plume stars]  is a photometrically clean sample  of genuine CMa stars,
un-contaminated by Galactic disk stars.  
This assumption however is in  complete contradiction with the  recent
finding of Carraro et al.  (\cite{carraro05}), who have shown that the
distribution of the blue plume population in the background of 30 open
clusters (in  the Third quadrant  and at $R_{\odot}\ge7$  kpc) follows
the expected pattern of the Norma-Cygnus (outer) spiral arm remarkably
well.
Thus  the  CMa over-density resides on  an  outer spiral  arm, and the
young blue plume population  (seen in various  CMa diagrams) is simply
not associated with it.
This is a  particularly important    issue  because it  explains   the
difference in the proper motion at the  CMa-center obtained by Dinescu
et   al.  (\cite{dinescu05}, using   MS   stars)  and  Momany  et  al.
(\cite{momany04b}, using  RGB stars) which in turn  brings us  back to
the situation where the proper motion of CMa is indistinguishable from
surrounding disk  stars\footnote{Note also  that in  calculating  $W$,
Dinescu et al.  (i) apply a mean  $V_{rad}$ value referring to the {\em
artificially produced $V_{rad}$  peak} at $109\pm4$  km/s by Martin et
al.  (\cite{martin04b}); (ii) assume a distance  of 8.1 kpc, different
from  the  most  recent   determination (7.2  kpc)   by Bellazzini  et
al. (\cite{luna05}); and (iii) combine the proper motion based on blue
plume stars with $V_{rad}$ values that are based on red clump stars.}.

Secondly, the fact that the negative $W$ of CMa is irreconcilable with
the expected signature  of the Galactic  warp (positive $W$) is simply
not  true, since  these   expectations are model-dependent.   We  will
demonstrate this  by using the same  cited  article of  Drimmel et al.
(\cite{drimmel00}), paying more attention to what these authors report
at the end of their article.

We first remind   the reader that  Drimmel et  al.  (\cite{drimmel00})
used Hipparcos  OB stars as tracers  of the Galactic warp.  Therefore,
it is particularly important to recall recent problems in the inferred
Hipparcos   parallaxes  of     O   stars.  Schr{\"   o}der     et  al.
(\cite{schroder04}) using a sample  of 153 stars, conclude that  their
absolute magnitude (calculated  from their apparent magnitude  and the
Hipparcos parallax) appear to  be much fainter  than expected, {\em by
up to 5 magnitudes}.   Such huge differences  were found to arise from
the distances   at which  O  stars  were  located,   and the level  of
precision of the parallax  measurements achieved by  Hipparcos.  Their
Figure~2  clearly shows that large  magnitude differences are expected
when using the  relatively uncertain Hipparcos parallaxes at distances
of already $\sim$1 kpc.

Accounting for this  fact,   we  reconsider the  work  of  Drimmel  et
al. (\cite{drimmel00}).  In their Figure~11,  they report a comparison
of  the measured  vertical    velocity   component ($W$)  {\it     vs}
Galactocentric distances.  At  $R_{GC}\simeq8.5$  kpc,   a
divergence   between  measured  $W$  values    of  distant OB    stars
(increasingly  negative  $W$ values)    and  the expected   observable
signature of the warp (increasingly positive  $W$ values) occurs.  
It  remains,  however, that  Drimmel  et al.  (\cite{drimmel00}, their
Section~7) investigate the effect of  bias, amplitude and precision on
the  inconsistency  between the observed    and the expected  warp $W$
values.   In their  Figure  12, the  authors  unambiguously  show that
``{\em negative  vertical  motions are  finally achieved}'',   if they
allow for: (i) a warp that has half the  amplitude with respect to the
one   they    derive;      (2)   a  warp     precessing     at   $-25$
km~s$^{-1}$~kpc$^{-1}$; and (iii) a $0.5$ magnitude error is added.

Now, leaving  aside the  excessively  high precision rates  which 
Drimmel et al. (\cite{drimmel00}) also  find questionable, it is a matter of
fact that  the remaining two  bias  strongly contribute in reproducing
negative vertical velocities.
On the one  hand,  the  results  presented  in   this paper (see    in
particular  Fig. 2  of Yusifov \cite{yusifov04})   show that the  warp
amplitude   as derived by Drimmel  et  al.  (\cite{drimmel00}) is very
high.
Indeed, at a Galactocentric distance of  10 kpc, Drimmel  et al.  note
that their  warp amplitude is more  than twice the gas warp amplitude,
and not  surprisingly they {\em halve it}.   In particular, the authors
note that the effect of misplacement of a star  to larger distances is
that its  measured relative vertical motion  will be smaller  than its
true relative vertical motion.

On  the other hand, there  is now enough evidence  to suggest that the
Hipparcos data  can have significantly higher   errors as suggested by
Schr{\" o}der et al.  [see also Soderblom et al.  (\cite{soderblom05})
on a recent confirmation of errors in the Hipparcos parallaxes already
at the Pleiades distance].   
Thus,  introducing a     0.5  mag  (as done    by    Drimmel  et   al.
\cite{drimmel00}) could have been seen as extreme at that time.
Nowadays, however,  with confirmed  errors of  more  than 2 magnitudes
(for the   same sample of   hot stars) at  much  closer distances, the
evidence suggests that {\em the expected warp signature  can be and is
compatible with negative vertical velocity}.

Since  Dinescu et al.   mistake  main sequence   stars belonging  to a
Galactic spiral  arm for genuine  CMa population, and  given the above
discussion on  the compatibility of  negative $W$ values with the warp
at  CMa location, we  conclude that the  proper motion of CMa does not
show any peculiar signature with respect to Galactic disk stars.

\section{The ``narrowness'' of  CMa main sequence}

Recently Mart{\'{\i}}nez-Delgado et al.  (\cite{delgado05})  presented
a  deep  wide-field $B$, ($B-R$) color-magnitude  diagram  of  the CMa
center and derived the line-of-sight extent (or depth) of CMa.
For this task, Mart{\'{\i}}nez-Delgado  et al.  estimated the observed
width  of the CMa main   sequence  ($\sigma_{\rm MS, total}^{B}$)   by
selecting stars in a   narrow color range,  $1.5\le~(B-R)~\le1.55$, at
$B\simeq22.1$.
Their  best fitting of $\sigma_{\rm   MS, total}^{B}$ (converted  into
kilo-parsec) yielded FWHM$=1.95$  kpc.   Moreover,  and for a   proper
comparison with other studied Local Group dwarf spheroidals, they also
estimated  the    line-of-sight   half-brightness radius  ($r_{1/2}$),
obtaining $\sim1$ kpc.

Mart{\'{\i}}nez-Delgado et al.  (\cite{delgado05}) noted that this size
is ``{\em significantly bigger than  that of several dwarf galaxies in
the Local    Group}'', e.g.  the $r_{1/2}$    of the Fornax  dwarf
spheroidal is   0.33 kpc  (see Table~4  of   Irwin    \& Hatzidimitriou
\cite{irwin95}). 
Nevertheless,   one  finds that    this rather  large  value  found by
Mart{\'{\i}}nez-Delgado et  al.  (\cite{delgado05}) is  often cited in
support   of  a ``narrow  extent   line-of-sight'' of  CMa  (Martin et
al. \cite{martin05}, Conn et al.  \cite{conn05}) or even evidence of a
``typical   size of  a     dwarf spheroidal galaxy''  (Bellazzini   et
al. \cite{luna05}).

In regards to this, and in light of recent finding of the presence of the
Norma-Cygnus  (outer) spiral arm   in the third  quadrant  (Carraro et
al. \cite{carraro05}),  we compare  the FWHM of   CMa  as obtained  by
Mart{\'{\i}}nez-Delgado et al.  (\cite{delgado05}) with typical values
for Galactic spiral arms.
Following Bronfman et al.  (\cite{bronfman}), we  estimate the FWHM of
the  most distant spiral  arm   at  $R/R_{\rm o}=1.3$, where   $R_{\rm
o}=8.5$ kpc.  Relying on their Fig.8,  we estimate a FWHM of $\sim1.7$
kpc,  very much in  accordance with the Mart{\'{\i}}nez-Delgado et al.
(\cite{delgado05}) 1.95 kpc value for CMa.  Indeed, one must also bear
in mind that  an outer  spiral arm would   also be {\em  more flared},
thereby a FWHM of $\sim1.7$ kpc sets only a lower limit.
One last piece  of evidence that the  reported CMa FWHM is  compatible
with a spiral arm feature comes from the study by McClure-Griffiths et
al.  (\cite{mcclure05}).   The detected  \hi\ arm  in the  Fourth  and
Third quadrants appeared to  be approximately 1 to  2 kpc thick  along
the line of sight.

We therefore      conclude   that the  CMa   depth    as  measured  by
Mart{\'{\i}}nez-Delgado  et al.  (\cite{delgado05}) is more compatible
with a distant spiral arm rather than a dwarf spheroidal.

\section{CMa and the Orion arm connection}

Recently, Moitinho  et al.   (\cite{moitinho06}) used young ($\sim100$
Myr) open clusters  and field "blue  plume" population  to reconstruct
the  spiral structure in the Third  Galactic quadrant.  They confirmed
the detection of a structure interpreted as a natural extension of the
Norma-Cygnus    spiral arm.  Their    conclusion  (based   on  $UBVRI$
photometry of    OB objects) is   consistent  with our  finding  of an
extended MW disk (using 2MASS M-giants).  Moitinho et al.  went a step
further and proposed that the CMa  M-giant over-density is simply the
result of looking  along the extension of  the local Orion arm  in the
Third quadrant.

In     regards  to     this,    we note     that     the  Moitinho  et
al. (\cite{moitinho06}) and  Carraro et al.  (\cite{carraro05}) papers
have been valuable in clearing  the connection between the blue  plume
population (erroneously  attributed to  CMa by Mart{\'{\i}}nez-Delgado
et al. \cite{delgado05}, Bellazzini  et al. \cite{luna04} and  Dinescu
et   al.  \cite{dinescu05}) and  both  the  red   clump  and red giant
populations    (that   first brought to the     detection   of the CMa
over-density).  
Nevertheless, the Moitinho  et   al. analysis does  not   address, nor
preclude, the    very    occurrence  of  the     CMa   over-density at
$b\sim-8^{\circ}$.   
Indeed,  the  detection of a spiral  arm  composed by {\em young} blue
objects (best  identified  in $UBV$ photometry)  is not  sufficient to
affirm/conclude a Galactic nature for the un-ambiguous presence of the
{\em older}  ($2-10$ Gyr) CMa M-giant  over-density  (best seen in the
2MASS infrared photometry).
The asymmetry in the distribution of the MW disk stars (Galactic warp)
remains {\em the}   explanation for  the  observed asymmetric   number
density  at     $b\sim-8^{\circ}$ and  $b\sim+8^{\circ}$.    The  warp
signature   is   evident   in     Fig.~2     of Moitinho     et    al.
(\cite{moitinho06}).

\end{document}